%% file: root.tex
\documentclass[letterpaper,fleqn,authoryear]{cas-sc}

\usepackage[numbers]{natbib}

\def\tsc#1{\csdef{#1}{\textsc{\lowercase{#1}}\xspace}}
\tsc{WGM}
\tsc{QE}

\newcommand{\mcal}[1]{\mathcal{#1}}

\newcommand{\mbf}[1]{\mathbf{#1}}

\DeclareMathOperator*{\argmin}{arg\,min} 
\usepackage{booktabs}
\usepackage{multirow}
\usepackage{amsthm}

\usepackage{algorithm}
\usepackage{algpseudocode}

\begin{document}
\let\WriteBookmarks\relax
\def\floatpagepagefraction{1}
\def\textpagefraction{.001}

\shorttitle{\rmfamily Optimal Experimental Design via Active Learning for Machine Learning-Based Building Energy Modeling}    

\shortauthors{NT Nguyen, TX Nghiem}  

\title [mode = title]{Active Learning for Optimal Experimental Design in Machine Learning-Based Building Energy System Identification}



%

\author[]{Nam T. Nguyen}[orcid=0009-0003-6914-1303]



\ead{nam.nguyen2@ucf.edu}

\ead[url]{https://namnguyenee2.github.io/}


\affiliation[]{organization={Department of Electrical and Computer Engineering, College of Engineering and Computer Science},
            addressline={ University of Central Florida}, 
            city={Orlando},
            postcode={ FL 32822}, 
            country={United States of America}}

\author[]{Truong X. Nghiem}[orcid=0000-0003-4841-3800]


\ead{truong.nghiem@ucf.edu}

\ead[url]{https://www.ece.ucf.edu/person/truong-nghiem/}





\nonumnote{This material is based upon work supported by the National Science Foundation under Awards No. 2513096 and 2514584.}

\begin{abstract}
Machine learning (ML) techniques have been commonly adopted to identify the dynamics of building energy systems (BESs), owing to their flexibility relative to first-principles, physics-based modeling approaches.
Beyond the choice of ML architecture, the quality of the training data plays an essential role in the resulting model performance.
Optimal experimental design (OED), realized in this work through active learning (AL), determines which experiments to conduct in order to collect informative data, rather than relying on standard approaches such as uniformly random sampling.
This paper proposes a systematic comparison of OED via AL for building energy system identification, with a particular focus on HVAC thermal dynamics.
We investigate fourteen AL techniques across two ML model classes, namely a deterministic feedforward neural network and a stochastic Gaussian process, and classify these techniques into four categories: data space, uncertainty, information gain, and model change.
To examine the AL algorithms under realistic conditions, we implement and evaluate them on the high-fidelity building simulator BOPTEST.
The results, reported as the root mean square error across multiple test scenarios with varying initial dataset sizes and control input constraints, show that AL-based models generally outperform models trained via passive learning (PL) with uniformly random control inputs, achieving error reductions of up to 54\%, although the magnitude and consistency of this improvement vary across acquisition functions and operating regimes.
\end{abstract}



\begin{keywords}
 \sep Active Learning
 \sep Optimal experimental design
 \sep Machine learning
 \sep Data-space sampling
 \sep Uncertainty sampling
 \sep Information-gain sampling
 \sep Model-change sampling
 \sep Building energy systems
\end{keywords}

\maketitle
\input{introduction}
\input{landscape}
\input{deterministic}
\input{stochastic}

\input{methodology}
\input{results}
\input{Conclusion}



\bibliographystyle{cas-model2-names}
\bibliography{references}


\end{document}

%% file: introduction.tex
\section{Introduction}
Optimizing energy consumption in building energy systems (BESs) is a critical concern, as building energy consumption is projected to account for more than 60\% of global electrical energy consumption in Society Year 2025-2026, according to ASHRAE’s Public Policy Priorities report \cite{ashraepolicy2026}.
In particular, the rapid growth of data centers in the AI era demands intelligent building control systems to avoid unnecessary energy waste \cite{10089168}.
Optimal control techniques, such as model predictive control \cite{DRGONA2020190} and reinforcement learning \cite{8356086}, have demonstrated the capability to reduce energy cost by up to 27.4\% in over 48 buildings.
However, these control techniques require an accurate predictive or dynamics model to achieve optimal control performance as well as guarantee the physics constraints of the systems.
Machine learning (ML) models have become increasingly popular for system modeling tasks due to their flexibility and accuracy in identifying BESs \cite{seyedzadeh2018machine, 11346501}, relying only on data collected from real-world or simulated environments.
However, data quality plays one of the most significant roles in determining the accuracy of the learned dynamics model.
In many cases, the training data for ML models is collected through randomly designed experiments using Pseudo-Random Binary Sequences (PRBS) \cite{9200539}, white-noise \cite{keesman2011system}, or uniform random signals \cite{11346501} for the control inputs.
For the sake of identifying nonlinear systems, these random excitation signals may require collecting a massive amount of data to accurately estimate the system dynamics \cite{JMLR:v23:20-807}.

To collect an informative dataset, optimal experimental design (OED) techniques have been developed in \cite{smucker2018optimal, atkinson2007optimum}, in which data are collected via information gain functions so that the resulting model achieves a lower estimation error than under random experiments. 
On the other hand, Active Learning (AL) schemes used in ML models follow the same underlying mechanism as OED.
Indeed, AL methods are mostly used to query which unlabeled data points to label, via acquisition functions \cite{settles2009active}.
The scope of this paper is to leverage these acquisition functions, which may equivalently be regarded as the information functions of OED, to design optimal experiments for accurately estimating BESs.
Within AL, two main paradigms are distinguished: pool-based AL and population-based AL \cite{8475012}.
In this paper, we study a dynamical system in a regression setting and collect time-series data, and therefore focus on population-based AL techniques.

Within the building energy domain, AL has been applied predominantly to static or short-horizon prediction tasks rather than to online dynamical system identification, and in every case a single acquisition function already established in the general AL literature is adopted, rather than situated within the broader taxonomy reviewed in Section~\ref{sec:landscape}.
\cite{mahmood2024active} applied a standard uncertainty sampling query strategy, wrapped around eight regressors including random forest and CatBoost, to predict cooling and heating loads in green buildings, reporting accuracy improvements over passively trained ensembles; however, the underlying dataset is a fixed set of 768 Ecotect-simulated building shapes rather than a sequentially collected trajectory, so this comparison addresses pool-based sampling from a static dataset rather than the sequential, dynamics-constrained control input design problem considered in this paper.
\cite{yang2025application} likewise adopted the classical maximum-uncertainty query function, combining it with random forest and CatBoost regressors to predict HVAC fan power consumption, but contrasted it only against random sampling within a two-by-two model and sampling design, rather than against the broader family of acquisition functions considered in this paper.
\cite{ZHANG2021111436} and \cite{ZHANG2021111026} both adopt the same expected error reduction acquisition function, originally developed for text classification and adapted here with a block design module that passively accounts for weather disturbances, to generate informative training data for data predictive control and short-term forecasting models, respectively; while this block design module is a genuine methodological contribution for handling uncontrollable disturbances, both studies evaluate only this single acquisition criterion and do not compare it against the uncertainty, information gain, or model change alternatives available for the same regression task.

AL has also been applied to classification and personalization tasks within heating, ventilation, and air conditioning (HVAC) operation, and to a single building component, rather than to the regression-based system identification problem considered in this paper.
\cite{FAN2024122356} compared five existing acquisition functions, namely least confidence, margin sampling, entropy, and two query-by-committee variants based on Monte Carlo dropout and Bayesian flipout, for HVAC fault diagnosis, reducing labeling costs by approximately fifty percent; however, all five acquisition functions are formulated for multiclass classification through softmax class probabilities and therefore do not transfer directly to the continuous-valued regression setting addressed by the uncertainty and model change categories in this paper.
\cite{tekler2023enhancing} employed the query-by-committee strategy to personalize thermal comfort models from occupant feedback, reporting a 31\% reduction in labeling effort together with a 1.3\% increase in energy savings, yet evaluated this single acquisition function only against a fully labeled baseline rather than against competing AL categories.
\cite{LI2024100189} is the only building energy study reviewed here that proposes a genuinely new acquisition function rather than adopting an established one, introducing a Mahalanobis distance and information density-based sampling strategy that deliberately constrains exploration to remain close to the current operating point for operational safety, alongside a comparison with four existing methods, namely query-by-committee, expected error reduction, variance reduction, and maximizing expected model change, on cooling tower data from a real high-rise building.
Taken together, there is a lack of a systematic study that comprehensively classifies and investigates different AL techniques in BESs, which is essential for future studies to select appropriate AL classifications and benchmarks before proposing new AL techniques for BESs.
Moreover, they usually work with a single ML type, either a deterministic or a stochastic model.

Motivated by these gaps, this paper proposes a systematic categorization of AL schemes for building energy system identification, organized into four categories, namely data space, uncertainty, information gain, and model change, in contrast to the single acquisition function typically adopted in prior building energy studies.
The rationale for this categorization, together with a detailed literature review of each category, is presented in Section~\ref{sec:landscape}.
These four categories are evaluated across two ML model classes, namely a feedforward neural network and a Gaussian process, corresponding to a deterministic model and a stochastic model, respectively, yielding a total of fourteen AL algorithms investigated in this study.
Every algorithm is benchmarked against passive learning (PL), in which the control inputs are sampled uniformly at random, a comparison that has not been conducted comprehensively for HVAC systems in the existing literature.

For a comprehensive comparison, all AL algorithms are examined on the Building Optimization Testing Framework (BOPTEST) \cite{Blum03092021}, a high-fidelity simulator that uses the Modelica Buildings library \cite{Wetter04072014} to model realistic HVAC dynamics derived from real-world systems.
The results demonstrate that, for both the NN and GP models, AL-based methods generally achieve better accuracy than PL with uniformly random control inputs, although the magnitude and consistency of this improvement vary across acquisition functions and test scenarios.

The main contributions of this paper are summarized as follows.
\begin{itemize}
    \vspace{-5pt}
    \item A systematic taxonomy of fourteen active learning  techniques for building energy system identification, organized into four categories, namely data space, uncertainty, information gain, and model change, and formally derived for both a deterministic ML model and a stochastic ML model.
    \vspace{-5pt}
    \item An implementation and evaluation of all fourteen techniques on the high-fidelity BOPTEST simulator, in contrast to prior building energy active learning studies that rely on static historical datasets, ad hoc virtual testbeds, or a single real building.
     \vspace{-5pt}
    \item A comprehensive empirical comparison across six test scenarios, spanning both model classes and a range of initial dataset sizes and control input ramp constraints, rather than a single fixed experimental configuration.
\end{itemize}

The remainder of this paper is organized as follows.
Section~\ref{sec:prelim} formalizes PL and AL under a unified online algorithmic structure, and presents the deterministic neural network and stochastic Gaussian process models considered throughout this paper.
Section~\ref{sec:landscape} surveys the broader literature on AL methods and organizes the fourteen techniques investigated in this paper into the four-category taxonomy introduced above.
Section~\ref{sec:deterministic} and Section~\ref{sec:stochastic_al} formally derive the acquisition functions, together with their respective advantages and limitations, for the deterministic neural network and the stochastic Gaussian process model, respectively.
Section~\ref{sec:method} describes the BOPTEST simulation setup for HVAC dynamics.
Section~\ref{sec:results} reports and discusses the resulting simulation results across all six test scenarios.
Finally, Section~\ref{sec:conclusion} concludes the paper and outlines future work.

%% file: landscape.tex
\section{Preliminaries}
\label{sec:prelim}

In terms of working with data-driven models, we consider a general discrete-time dynamical system, which is used to represent thermal dynamics in HVAC systems later in this paper, with the assumption that the system is fully observable
\begin{align}
    \mathbf{x}_{k+1} = f(\mathbf{x}_k, \mathbf{u}_k), \label{eq:ODE}
\end{align}
where $\mathbf{x}_k \in \mathbb{R}^{n_x}$ and $\mathbf{u}_k \in \mathbb{R}^{n_u}$ denote the system state vector and the control input vector at discrete time step $k$, respectively, and $f: \mathbb{R}^{n_x} \times \mathbb{R}^{n_u} \rightarrow \mathbb{R}^{n_x}$ is an unknown nonlinear and non-convex function governing the system dynamics. 
The central objective of this work is to learn an accurate neural network (NN) approximation of $f$ from a dataset $\mcal{D} = \left\{\left[\mathbf{x}_i, \mathbf{u}_i\right]; \mathbf{x}_{i+1}\right\}_{i=0}^{N}$, where the input to the NN is the concatenation of the state vector $\mathbf{x}_i$ and the control input vector $\mathbf{u}_i$, and the corresponding regression target is the successor state $\mathbf{x}_{i+1}$. 
Since acquiring labeled observations of the system requires executing physical experiments, which is both time-consuming and resource-intensive, it is essential that the collected dataset $\mcal{D}$ is maximally informative rather than constructed through naive random sampling.

\textbf{Notation}: Let $\mcal{Z} = \mcal{X} \times \mcal{U}$ denote the joint space of the concatenated state-input pairs, where $\mcal{Z}_c \subset \mcal{Z}$ is the subspace of already collected (labeled) samples and $\mcal{Z}_u \subset \mcal{Z}$ is the subspace of candidate (unlabeled) samples, such that $\mcal{Z}_c \cup \mcal{Z}_u = \mcal{Z}$. 
For each sample, we define the concatenated vector $\mathbf{z} = [\mathbf{x}, \mathbf{u}] \in \mcal{Z}$, where $\mathbf{x} \in \mcal{X}$ and $\mathbf{u} \in \mcal{U}$.

\subsection{Passive Learning and Active Learning}

The most straightforward strategy for constructing the dataset $\mcal{D}$ during online operation is passive learning (PL), in which the control input $\mathbf{u}_k$ applied at every time step $k$ is drawn from a random distribution over the admissible input space $\mcal{U}$, independently of the current system state, the observations collected so far, and the model trained on them \cite{Bao02072024}. 
Because PL disregards any notion of which experiments are most informative for identifying the unknown dynamics $f$, it is adopted throughout this study as the natural baseline against which every OED-based sampling strategy is compared, rather than as a competitive design method in its own right. Algorithm~\ref{alg:passive} formalizes this online PL procedure: an initial model is trained on the available data, after which randomly sampled control inputs are repeatedly applied to the system, the resulting state observations are appended to $\mcal{D}$, and the model is periodically retrained as additional data accumulate.

\begin{algorithm}[t]
    \caption{Passive Learning with Online Updating}
    \label{alg:passive}
    \begin{algorithmic}
    \small
    \Require Initialize the training data $\mcal{D}$, initialize ML model parameters $\theta$, number of experiments $N_\text{budget}$.
    \Ensure Optimal parameters $\theta^*$ 
    \State Train $\theta_0 \leftarrow \argmin_{\theta} \mcal{L}(\theta; \mcal{D})$.
    \For $\;\;i \!: 1 \rightarrow N_\text{budget}$
        \State $\mbf{u}_i \sim \mcal{U}$ (randomly) with current state $\mbf{x}_{i}$
        \State Given $[\mbf{x}_i, \mbf{u}_i]$, we collect $\mbf{x}_{i+1}$ via \eqref{eq:ODE}.
        \State Update the dataset $\mcal{D} = \mcal{D} \cup \{ [\mbf{x}_i, \mbf{u}_i], \mbf{x}_{i+1}\}$.
        \State Retrain the ML model $\theta_i \leftarrow \argmin_{\theta} \mcal{L}(\theta; \mcal{D}) $
    \EndFor
    \State Save optimal parameters $\theta^* \leftarrow \theta_{N_\text{budget}}$.
    \end{algorithmic}
\end{algorithm}

\begin{algorithm}[t]
    \caption{Active Learning with Online Updating}
    \label{alg:active}
    \begin{algorithmic}
    \small
    \Require Initialize the training data $\mcal{D}$, initialize ML model parameters $\theta$, number of experiments $N_\text{budget}$.
    \Ensure Optimal parameters $\theta^*$ 
    \State Train $\theta_0 \leftarrow \argmin_{\theta} \mcal{L}(\theta; \mcal{D})$.
    \For $\;\;i \!: 1 \rightarrow N_\text{budget}$
        \State $\mathbf{u}_i^* = \underset{\mathbf{u} \in \mcal{U}}{\arg\max} \ \alpha(\mathbf{x}_i, \mathbf{u})$
        \State Given $[\mbf{x}_i, \mbf{u}_i]$, we collect $\mbf{x}_{i+1}$ via \eqref{eq:ODE}.
        \State Update the dataset $\mcal{D} = \mcal{D} \cup \{ [\mbf{x}_i, \mbf{u}_i], \mbf{x}_{i+1}\}$.
        \State Retrain the ML model $\theta_i \leftarrow \argmin_{\theta} \mcal{L}(\theta; \mcal{D}) $
    \EndFor
    \State Save optimal parameters $\theta^* \leftarrow \theta_{N_\text{budget}}$.
    \end{algorithmic}
\end{algorithm}
Although PL is simple to implement and requires no additional computation for selecting $\mathbf{u}_k$, it allocates experiments randomly across the admissible space regardless of where the current model is most uncertain or stands to benefit the most from additional data, which generally makes inefficient use of a limited experimental budget \cite{8759089}. 

To this end, active learning (AL) is adopted as a principled framework for sequentially designing the control input $\mathbf{u}_k$ at each time step $k$ such that the resulting observations contribute the greatest possible information gain toward accurately learning $f$.
Specifically, at each step $k$, given the current state $\mathbf{x}_k$, the AL algorithm determines the optimal control input $\mathbf{u}_k^*$ by solving the following population-based acquisition optimization problem
\begin{align}
    \mathbf{u}_k^* = \underset{\mathbf{u} \in \mcal{U}}{\arg\max} \ \alpha(\mathbf{x}_k, \mathbf{u}), \label{eq:opti_active}
\end{align}
where $\alpha: \mathbb{R}^{n_x} \times \mathbb{R}^{n_u} \rightarrow \mathbb{R}$ is the acquisition function that quantifies the expected informativeness of a candidate input $\mathbf{u}$ given the current state $\mathbf{x}_k$ and the dataset $\mcal{D}$ collected thus far.
Algorithm~\ref{alg:active} formalizes the resulting online AL procedure, which follows exactly the same structure as Algorithm~\ref{alg:passive} except that the randomly sampled input is replaced, at every step, by the input $\mathbf{u}_k^*$ that solves the acquisition optimization problem in \eqref{eq:opti_active}, after which the dataset is updated and the model is retrained as in the passive case.

\subsection{Deterministic and Stochastic Model}
\label{sec:models}
The central objective of this work is to learn a data-driven approximation of the unknown dynamics $f$ using either a deterministic or a stochastic ML model.
Both model classes take the concatenated state-input vector $\mathbf{z}_k = [\mathbf{x}_k, \mathbf{u}_k] \in \mathbb{R}^{n_z}$, where $n_z = n_x + n_u$, as input and produce a prediction of the successor state $\mathbf{x}_{k+1} \in \mathbb{R}^{n_x}$ as output.
Their key distinction lies in the nature of the output: a deterministic model yields a point prediction, whereas a stochastic model yields a full predictive distribution, providing a native measure of uncertainty that is directly exploited by the active learning acquisition functions described in Section~\ref{sec:stochastic_al}.

\subsubsection{Feedforward Neural Network}
\label{sec:NN}
A feedforward neural network (NN) is a parametric model that approximates the unknown function $f: \mathbb{R}^{n_z} \rightarrow \mathbb{R}^{n_x}$ by composing a sequence of affine transformations and nonlinear activation functions.

Let $L$ denote the number of hidden layers and $n_h$ denote the number of neurons per hidden layer, with $n_{h,0} = n_z$ and $n_{h,\ell} = n_h$ for $\ell = 1, \ldots, L$.
The NN is defined by the following recursive computation:
\begin{align*}
    \mathbf{h}_0 &= \mathbf{z}_k, \\
    \mathbf{h}_\ell &= \varphi\!\left(\mathbf{W}_\ell\, \mathbf{h}_{\ell-1} + \mathbf{b}_\ell\right), \quad \ell = 1, \ldots, L, \\
    \hat{f}_\text{NN}(\mathbf{z}_k;\, {\theta}) &= \mathbf{W}_{L+1}\, \mathbf{h}_L + \mathbf{b}_{L+1},
\end{align*}
where $\mathbf{W}_\ell \in \mathbb{R}^{n_{h,\ell} \times n_{h,\ell-1}}$ and $\mathbf{b}_\ell \in \mathbb{R}^{n_{h,\ell}}$ are the weight matrix and bias vector of the $\ell$-th layer, respectively, and $\varphi: \mathbb{R} \rightarrow \mathbb{R}$ is a nonlinear activation function applied element-wise.
The final layer applies a linear transformation to produce the output $\hat{f}_\text{NN} \in \mathbb{R}^{n_x}$, and the complete set of trainable parameters is ${\theta} = \{\mathbf{W}_\ell, \mathbf{b}_\ell\}_{\ell=1}^{L+1} \in \mathbb{R}^{n_\theta}$.

Given a dataset $\mcal{D}_k = \{[\mathbf{x}_i, \mathbf{u}_i]; \mathbf{x}_{i+1}\}_{i=0}^{k-1}$ of $k$ labeled samples collected up to step $k$, the parameters ${\theta}$ are obtained by minimizing the regularized mean squared error (MSE) loss:
\begin{align}
    \mcal{L}({\theta}; \mcal{D}) = \frac{1}{k} \sum_{i=0}^{k-1} \left\|\mathbf{x}_{i+1} - \hat{f}_\text{NN}(\mathbf{z}_i;\, {\theta})\right\|_2^2 + \lambda\, \|{\theta}\|_2^2, \label{eq:loss_NN}
\end{align}
where $\lambda \geq 0$ is the L2 regularization coefficient that prevents overfitting by penalizing large parameter magnitudes, which is particularly important in the early stages of active learning when the available dataset is small.

At test time, the trained NN produces a single deterministic point estimate of the successor state through a single forward pass
\begin{equation*}
    \hat{\mathbf{x}}_{k+1} = \hat{f}_\text{NN}(\mathbf{x}_k, \mathbf{u}_k;\, {\theta}).
\end{equation*}
Unlike the GP model (Section~\ref{sec:gp}), the NN provides no native measure of predictive uncertainty.
Epistemic uncertainty must therefore be externally quantified through ensemble-based or stochastic inference techniques, as described in Sections~\ref{sec:uncertainty} and~\ref{sec:model_change}.

\subsubsection{Gaussian Process}
\label{sec:gp}

A Gaussian process (GP) is a non-parametric probabilistic model that places a prior distribution over functions, enabling simultaneous prediction of a mean response and a calibrated measure of predictive uncertainty from a single fitted model.
In this work, the GP models the unknown dynamics as
\begin{equation*}
    y_k = f(\mathbf{z}_k) + \varepsilon_k, \quad \varepsilon_k \sim \mcal{N}(0,\, \sigma_n^2),
\end{equation*}
where $y_k \in \mathbb{R}$ is a scalar output observation, $f: \mathbb{R}^{n_z} \rightarrow \mathbb{R}$ is the latent function to be inferred, and $\varepsilon_k$ is independent Gaussian observation noise with variance $\sigma_n^2 > 0$.
Since the successor state $\mathbf{x}_{k+1} \in \mathbb{R}^{n_x}$ has $n_x$ components, a set of $n_x$ independent GPs is trained in parallel, one per output dimension.
The following development describes a single GP for one output dimension without loss of generality.

A GP prior over $f$ is fully specified by a mean function $m: \mathbb{R}^{n_z} \rightarrow \mathbb{R}$ and a positive-definite covariance (kernel) function $k: \mathbb{R}^{n_z} \times \mathbb{R}^{n_z} \rightarrow \mathbb{R}_{>0}$, denoted:
\begin{equation*}
    f(\mathbf{z}) \sim \mcal{GP}\!\left(m(\mathbf{z}),\; k(\mathbf{z}, \mathbf{z}')\right).
\end{equation*}
A zero mean function $m(\mathbf{z}) = 0$ is adopted throughout this work, as is standard for centered regression tasks.
The radial basis function (RBF) kernel, also referred to as the squared exponential kernel, is employed:
\begin{equation}
    k(\mathbf{z}, \mathbf{z}') = \sigma_f^2\, \exp\!\left(-\frac{\left\|\mathbf{z} - \mathbf{z}'\right\|_2^2}{2\ell^2}\right), \label{eq:rbf}
\end{equation}
where $\sigma_f^2 > 0$ is the signal variance and $\ell > 0$ is the length-scale parameter governing the smoothness of $f$.
The complete hyperparameter vector is ${\theta} = \{\sigma_f^2,\, \ell,\, \sigma_n^2\}$.

Given the training dataset $\mcal{D}_k = \{\mathbf{z}_i, y_i\}_{i=0}^{k-1}$ of $k$ observations, let $\mathbf{Z}_k = [\mathbf{z}_0, \ldots, \mathbf{z}_{k-1}]^\top \in \mathbb{R}^{k \times n_z}$ denote the matrix of training inputs and $\mathbf{y}_k = [y_0, \ldots, y_{k-1}]^\top \in \mathbb{R}^k$ the corresponding output vector.
The kernel matrix $\mathbf{K}_k \in \mathbb{R}^{k \times k}$ has entries $[\mathbf{K}_k]_{ij} = k(\mathbf{z}_i, \mathbf{z}_j)$, and the noisy kernel matrix is $\mathbf{K}_{y,k} = \mathbf{K}_k + \sigma_n^2\, \mathbf{I}_k \in \mathbb{R}^{k \times k}$, where $\mathbf{I}_k$ denotes the $k \times k$ identity matrix.

The hyperparameters ${\theta}$ are estimated by minimizing the negative log marginal likelihood of the training outputs
\begin{align}
    \mcal{L}_\text{GP}(\theta; \mcal{D}) &= -\log p(\mathbf{y}_k \mid \mathbf{Z}_k,\, {\theta}) \notag\\
    &= \frac{1}{2}\,\mathbf{y}_k^\top \mathbf{K}_{y,k}^{-1} \mathbf{y}_k + \frac{1}{2}\log\det\!\left(\mathbf{K}_{y,k}\right) + \frac{k}{2}\log(2\pi). \label{eq:loss_GP}
\end{align}
This objective automatically balances data fit (first term) against model complexity (second term), providing a principled and validation-free criterion for kernel hyperparameter selection.

Given a test input $\mathbf{z}_* = [\mathbf{x}_k, \mathbf{u}_k]$, the GP posterior predictive distribution conditioned on $\mcal{D}_k$ is Gaussian:
\begin{equation*}
    p(y_* \mid \mathbf{z}_*,\, \mcal{D}_k,\, {\theta}_k) = \mcal{N}\!\left(\mu_\text{GP}(\mathbf{z}_*;\, {\theta}_k),\; \sigma_\text{GP}^2(\mathbf{z}_*;\, {\theta}_k)\right),
\end{equation*}
with posterior predictive mean and variance
\begin{subequations}
\begin{align}
\mu_\text{GP}(\mathbf{z}_*;\, {\theta}_k) &= \mathbf{k}_*^\top\, \mathbf{K}_{y,k}^{-1}\, \mathbf{y}_k, \label{eq:gp_mean} \\
\sigma_\text{GP}^2(\mathbf{z}_*;\, {\theta}_k) &= k(\mathbf{z}_*, \mathbf{z}_*) - \mathbf{k}_*^\top\, \mathbf{K}_{y,k}^{-1}\, \mathbf{k}_*, \label{eq:gp_var}
\end{align}
\end{subequations}

where $\mathbf{k}_* = [k(\mathbf{z}_*, \mathbf{z}_0), \ldots, k(\mathbf{z}_*, \mathbf{z}_{k-1})]^\top \in \mathbb{R}^k$ is the vector of cross-covariances between the test point and all training inputs.
The mean $\mu_\text{GP}$ serves as the point prediction of the successor state component, while $\sigma_\text{GP}^2$ provides an analytically tractable, native measure of predictive uncertainty that directly reflects how well the test point is supported by the available training data.
This uncertainty estimate is the central quantity exploited by the stochastic active learning methods presented in Section~\ref{sec:stochastic_al}.

\section{Landscape of Active Learning}
\label{sec:landscape}
Building upon the discrete-time system model and the deterministic and stochastic learning architectures introduced in Section~\ref{sec:models}, this section surveys the active learning literature relevant to identifying dynamical systems such as the building thermal dynamics considered in this paper.
Active learning and its closely related counterpart, optimal experimental design (OED), have been studied extensively across the statistics, machine learning, and control communities.
Following this organizing principle, the methods reviewed below, and subsequently formalized in Sections~\ref{sec:data_space} through~\ref{sec:stochastic_al}, are classified into four categories: data space, uncertainty, information gain, and model change.
Table~\ref{tab:ode-techniques} summarizes the 14 specific techniques investigated in this study within this taxonomy, together with the associated model class, the acronym, and the section in which each acquisition function is formally derived.

\subsection{Data Space}

Data space methods select the next experiment using only the geometric structure of the already collected and candidate samples in the joint state-input space $\mcal{Z}$, without requiring any prediction or uncertainty estimate from the trained model.
The earliest instances of this idea originate from passive and active sampling strategies for static regression, in which the candidate point farthest from the existing labeled set is queried so as to maximize the diversity of the resulting design \cite{5694100}.
\cite{WU201990} formalized this maximin-distance principle as greedy sampling and further proposed variants that promote diversity in the output space in addition to the input space, while \cite{BEMPORAD2023275} extended the idea by combining a geometric exploration term with a model-based exploitation term through inverse distance weighting, yielding an acquisition function that balances coverage of the input space against regions of high prediction error.
\cite{10886678} later extended the inverse-distance-weighting framework from static regression to the identification of autoregressive dynamical models, demonstrating the applicability of data space sampling to the sequential, dynamics-driven setting considered in this paper.
Because data space methods depend only on the relative positions of samples in $\mcal{Z}$, they remain well defined even before any model has been trained, which makes them a natural choice during the cold-start phase of online learning, a property examined in detail for the Greedy Sampling and Inverse Distance Weighting methods adopted for the deterministic and stochastic models in Sections~\ref{sec:data_space} and~\ref{sec:data_space_gp}, respectively.

\subsection{Uncertainty}

Uncertainty sampling instead selects the next experiment by directly querying the candidate input at which the current model is least confident in its prediction, under the premise that reducing the largest sources of predictive uncertainty yields the most informative observations.
For models that provide only a point prediction, such as a standard feedforward neural network, predictive uncertainty must be approximated externally, for example through the disagreement among an ensemble of independently trained models in the query-by-committee framework \cite{buus2003sensitive} later adapted to regression by \cite{10.5555/1777942.1777965}, or through repeated stochastic forward passes with dropout retained at inference time, as proposed by \cite{pmlr-v48-gal16} under an approximate Bayesian interpretation of the dropout mechanism.
In contrast, models that natively output a predictive distribution, such as the Gaussian process employed in this study, admit uncertainty sampling directly from the analytically available posterior variance, as investigated by \cite{pmlr-v120-buisson-fenet20a} for actively learning Gaussian process dynamics, and extended to a globally integrated variance-reduction criterion by \cite{kontoudis2023closed} so as to account for the effect of a new observation across the entire admissible input space rather than only at the queried point.
These ensemble-based, dropout-based, and posterior-variance-based strategies are adopted in this study as Query-by-Committee and Monte Carlo Dropout for the neural network model, and as Maximize Variance and Integrated Variance Reduction for the Gaussian process model, formalized in Sections~\ref{sec:uncertainty} and~\ref{sec:uncertainty_gp}, respectively.

\subsection{Information gain}
Information gain methods adopt a more structural objective, selecting experiments that are most informative not about the predicted output at a single point but about the parameters or hyperparameters that govern the model itself.
This perspective is rooted in the classical theory of optimal experimental design, where the Fisher information matrix quantifies the amount of information that an experiment is expected to provide about an unknown parameter vector, and design criteria such as D-optimality, A-optimality, and E-optimality select experiments that maximize the determinant, trace, or minimum eigenvalue of this matrix, respectively \cite{smucker2018optimal}.
An information-theoretic counterpart to this classical framework formulates active learning instead through the mutual information, or equivalently the reduction in Shannon entropy, between the candidate observation and the unknown model parameters \cite{houlsby2011bayesianactivelearningclassification}, an idea originally developed for Bayesian classification and preference learning and later adapted to Gaussian process hyperparameter estimation by explicitly accounting for hyperparameter uncertainty through a Laplace approximation of the hyperparameter posterior \cite{3020751.3020776}, as adopted by \cite{8443729} for Gaussian-process-based building energy system identification.
Because both the Fisher information matrix and the Laplace approximation of the hyperparameter posterior are most naturally defined for models with an explicit parametric likelihood, such as the Gaussian process considered in this study, information gain sampling is investigated here only for the stochastic model, through the Fisher Information and Shannon Entropy methods formalized in Section~\ref{sec:info_gain_gp}.

Extending an analogous criterion to deep neural network parameters, for example, through Fisher-information-based gradient embeddings \cite{10.5555/3540261.3540944}, remains an active area of related research but is left outside the present scope due to the Fisher matrix computed for the NN model by not a direct way but needs to approximate the deterministic model to a stochastic model.
Therefore, we will investigate the information gain for the stochastic model only as it highlights the strength of the stochastic model.

\subsection{Model Change}

Model change methods select the experiment expected to induce the largest update to the trained model upon being observed, under the rationale that the samples producing the largest change in the model are the ones from which the model has the most to learn.
For gradient-based models such as the feedforward neural network, this principle is operationalized by estimating the norm of the loss gradient that a candidate observation would induce, first proposed for regression by \cite{Cai2013MEMC} through expected model change maximization using a bootstrap ensemble of models to approximate the unknown label distribution, and later made more computationally efficient by restricting the gradient computation to the final network layer, following the theoretical lower bound established by \cite{Ash2020Deep} for deep batch active learning.
For the Gaussian process model, the analogous notion of model change is instead expressed through Bayesian-optimization-style acquisition functions that explicitly balance exploitation of the current posterior mean against exploration of regions where the posterior is most likely to be revised by a new observation, namely the upper confidence bound criterion \cite{carpentier2011upper}, later extended to active learning of conditional mean embeddings by \cite{pmlr-v124-ray-chowdhury20a}, and the probability of improvement criterion \cite{7900034}, both reviewed in the broader context of Bayesian optimization by \cite{frazier2018tutorialbayesianoptimization}.
These four methods, namely Expected Model Change Maximization and Maximize Last Layer Change for the neural network model, and Upper Confidence Bound and Probability of Improvement for the Gaussian process model, are formalized in Sections~\ref{sec:model_change} and~\ref{sec:model_change_gp}, respectively.

Although every technique reviewed above instantiates the same general online active learning procedure formalized in Algorithm~\ref{alg:active}, the techniques differ from one another only through the specific definition of the acquisition function $\alpha(\cdot)$.
Therefore, in the sections that follow, the discussion focuses exclusively on deriving and analyzing the acquisition function of each method, with the understanding that every acquisition function considered in this paper is embedded within this same underlying algorithmic structure.

\textbf{Notation:} Let us denote $\alpha_{-}(\cdot)$ as the acquisition function with the subscript ``$-$''. 
Here, ``$-$'' is modified by following acronym of the considered method illustrated in Table \ref{tab:ode-techniques}.
For example, $\alpha_\text{mv}(\cdot)$ is the acquisition function of the maximizing variance sampling method.

\begin{table*}[t]
\centering
\normalsize
\caption{\rmfamily \normalsize Optimal experimental design (OED) techniques investigated in this study, grouped by category and model type.}
\vspace{0.2cm}
\label{tab:ode-techniques}
\renewcommand{\arraystretch}{1.02}
\setlength{\tabcolsep}{5pt}
\rmfamily
\normalsize
\makebox[\textwidth][c]{\resizebox{\dimexpr\textwidth\relax}{!}{\begin{tabular}{@{} l l c c l c c @{}}
\toprule
 & \multicolumn{3}{c}{\textbf{Deterministic model (NN)}} & \multicolumn{3}{c}{\textbf{Stochastic model (GP)}} \\
\cmidrule(lr){2-4} \cmidrule(lr){5-7}
\textbf{Category} & \textbf{Method} & \textbf{Acronym} & \textbf{Section} & \textbf{Method} & \textbf{Acronym} & \textbf{Section} \\
\midrule
\multirow{2}{*}{Data space}
    & Greedy sampling            & GS$_\text{NN}$  &  \ref{sec:greedy} & Greedy sampling            & GS$_\text{GP}$   & \ref{sec:gs_gp} \\
    & Inverse distance weighting & IWD$_\text{NN}$  & \ref{sec:idw} & Inverse distance weighting & IWD$_\text{GP}$ & \ref{sec:idw_gp}\\
\midrule
\multirow{2}{*}{Uncertainty}
    & Query-by-committee   & QBC & \ref{sec:qbc} & Maximize variance             & MV  & \ref{sec:mv}\\
    & Monte Carlo dropout  & MCD & \ref{sec:mcd} & Integrated variance reduction & IVR & \ref{sec:ivr} \\
\midrule
\multirow{2}{*}{Information gain}
    & & & &  Fisher information & FI & \ref{sec:fi} \\
    & & & & Shannon entropy &  SE & \ref{sec:se}\\
\midrule
\multirow{2}{*}{Model change}
    & Expected model change maximization & EMCM & \ref{sec:EMCM} & Upper confidence bound & UCB & \ref{sec:ucb} \\
    & Maximize last layer model change & MLLC & \ref{sec:mllc} & Probability of improvement & PI  & \ref{sec:pi}\\
\bottomrule
\end{tabular}}}
\end{table*}

%% file: deterministic.tex
\section{Active Learning for Deterministic Models}
\label{sec:deterministic}
This section formally derives the acquisition function of each of the six AL methods investigated for the deterministic NN model, organized into three categories: data space (Section~\ref{sec:data_space}), uncertainty sampling (Section~\ref{sec:uncertainty}), and expected model change (Section~\ref{sec:model_change}).
For each method, the corresponding advantages and limitations are analyzed from the perspective of designing efficient experiments for building energy system identification.


\subsection{Data Space}
\label{sec:data_space}

The data space approach constructs optimal experiments by relying solely on the geometric structure of the joint input space, the output space, or both.
Greedy sampling exploits only the input space, whereas Inverse Distance Weighting draws on both the input and output spaces.

\subsubsection{Greedy Sampling}
\label{sec:greedy}
Greedy sampling (GS$_\text{NN}$) is a model-independent active learning strategy, originally proposed as passive sampling for regression by \cite{5694100} and further developed for active learning regression by \cite{WU201990}.
The core principle of GS$_\text{NN}$ is to promote maximal spatial diversity across the joint state-input space $\mcal{Z}$ by sequentially selecting the candidate sample that is farthest from all previously collected labeled samples, entirely based on geometric distances in the feature space without any dependence on the current NN model. 

Denote the $T$ labeled samples as $\mbf{z}^{(c)} \in \mcal{Z}_c$ for $c = 1, \dots, C$, and the candidate samples as $\mbf{z}^{(u)} \in \mcal{Z}_u$ for $u = 1, \dots, U$. The pairwise Euclidean distance between a labeled sample $\mbf{z}^{(c)}$ and a candidate sample $\mbf{z}^{(u)}$ is defined as
\begin{equation*}
    d\!\left(\mbf{z}^{(c)}, \mbf{z}^{(u)}\right) = \left\|\mbf{z}^{(c)} - \mbf{z}^{(u)}\right\|_2,
\end{equation*}
and the minimum distance from the candidate sample $\mbf{z}^{(u)} = [\mbf{x}^{(u)}, \mbf{u}^{(u)}]$ to the entire labeled set $\mcal{Z}_c$ is defined as
\begin{equation*}
    d^{(c)}\!\left(\mbf{z}^{(u)}\right) = \min_{c = 1,\dots,C} \; d\!\left(\mbf{z}^{(c)}, \mbf{z}^{(u)}\right).
\end{equation*}
The optimal control input $\mbf{u}_k^*$ is then determined by maximizing this minimum distance over candidate control inputs
\begin{equation*}
    \mbf{u}_k^* = \underset{\mbf{u}^{(u)} \in \mcal{U}}{\arg\max} \; d^{(c)}\!\left(\mbf{x}_k, \mbf{u}^{(u)}\right)\text, \quad (\mbf{z}^{(u)}=[\mbf{x}_k, \mbf{u}^{(u)}])
\end{equation*}
where $\mbf{x}_k$ is the current system state at step $k$, which is fixed and not optimized.
The acquisition function is therefore defined directly from this maximin distance,
\begin{equation*}
    \alpha_\text{GS}(\mbf{u}_k) = d^{(c)}\!\left([\mbf{x}_k, \mbf{u}_k]\right).
\end{equation*}

The primary advantage of GS$_\text{NN}$ is its complete independence from the NN model, rendering it applicable at any stage of the active learning process, including the cold-start phase where no labeled data are yet available to train a reliable model. 
Furthermore, its computational cost per iteration scales only with the number of collected samples $C$, making it highly efficient for online deployment in building energy system identification. 
However, GS$_\text{NN}$ does not exploit any information from the model predictions or output observations, and consequently cannot distinguish between regions of the input space that are dynamically rich and those that are informationally redundant. 
This model-agnostic property can lead to suboptimal sample allocation in systems where the thermal dynamics vary significantly across different operating regions, since GS$_\text{NN}$ treats all regions of $\mcal{Z}$ as equally worth exploring.

\subsubsection{Inverse Distance Weighting}
\label{sec:idw}

The Inverse Distance Weighting method (IWD), formally introduced by \cite{BEMPORAD2023275}, combines data-space and model-based active learning principles for regression by constructing an acquisition function from two complementary terms: an exploitation term that targets regions of high model uncertainty, and a pure exploration term that promotes diversity by penalizing candidate inputs located near already-sampled regions of $\mcal{Z}$.
Unlike GS$_\text{NN}$, IWD explicitly incorporates the current NN model predictions into the acquisition function, thereby directing the sampling toward regions where the model is least reliable.
The method has also been extended to autoregressive models in \cite{10886678} for non-autonomous dynamical systems.
While working with NN models, we denote the IWD method as IWD$_\text{NN}$.
We assume that all variables are normalized using the min-max normalization method, such that each component of $\mbf{z} = [\mbf{x}, \mbf{u}]$ lies within the interval $[0, 1]$.

\noindent
Given the dataset $\mcal{D} = \left\{\left[\mbf{x}_i, \mbf{u}_i\right]; \mbf{x}_{i+1}\right\}_{i=0}^{k-1}$ collected up to step $k$, let $\mbf{z}_j = [\mbf{x}_j, \mbf{u}_j]$ denote the $j$-th collected sample in $\mcal{Z}$. The squared Euclidean distance between a candidate point $[\mbf{x}_k, \mbf{u}_k]$ and the $j$-th labeled sample $\mbf{z}_j$ is
\begin{equation*}
    d^2\!\left(\mbf{z}_j,\, [\mbf{x}_k, \mbf{u}_k]\right) = \|\mbf{z}_j - [\mbf{x}_k, \mbf{u}_k]\|_2^2.
\end{equation*}
The IDW weight function $w_j : \mcal{U} \rightarrow \mathbb{R}_{>0}$ is defined as
\begin{equation*}
    w_j(\mbf{u}_k) = \frac{e^{d^2\!\left(\mbf{z}_j,\, [\mbf{x}_k, \mbf{u}_k]\right)}}{d^2\!\left(\mbf{z}_j,\, [\mbf{x}_k, \mbf{u}_k]\right)}, 
\end{equation*}
which assigns larger weights to labeled samples located closer to the candidate point $[\mbf{x}_k, \mbf{u}_k]$. The normalized IDW coefficient function $v_j : \mcal{U} \rightarrow [0, 1]$ is then defined as
\begin{equation*}
    v_j(\mbf{u}_k) = \begin{cases}
        1 & \text{if } [\mbf{x}_k, \mbf{u}_k] = [\mbf{x}_j, \mbf{u}_j], \\[4pt]
        0 & \text{if } [\mbf{x}_k, \mbf{u}_k] = [\mbf{x}_h, \mbf{u}_h], \quad h \neq j, \\[4pt]
        \dfrac{w_j(\mbf{u}_k)}{\displaystyle\sum_{h=1}^{k} w_h(\mbf{u}_k)} & \text{otherwise.}
    \end{cases}
\end{equation*}
The IDW variance function $s^2 : \mcal{U} \rightarrow \mathbb{R}_{\geq 0}$, which serves as a proxy for the local prediction uncertainty of the NN model, is defined as
\begin{equation}
    s^2(\mbf{u}_k) = \sum_{j=1}^{k} v_j(\mbf{u}_k) \big\|\mbf{x}_{j} - \hat{f}_\text{NN}\!\left(\mbf{z}_{j-1};\, {\theta}_k\right) \big\|_2^2, \label{eq:IDW-NN-S}
\end{equation}
where $\hat{f}_\text{NN}(\mbf{z}_{j-1}; {\theta}_k)$ denotes the NN prediction of the successor state given the input $\mbf{z}_{j-1} = [\mbf{x}_{j-1}, \mbf{u}_{j-1}]$ and the current NN parameter vector ${\theta}_k$. The IDW exploration function $z : \mcal{U} \rightarrow [0, 1]$, which promotes diversity by returning higher values far from previously sampled inputs, is defined as
\begin{equation*}
    z(\mbf{u}_k) = \begin{cases}
        0 & \text{if } \mbf{u}_k \in \{\mbf{u}_0, \dots, \mbf{u}_{k-1}\}, \\[4pt]
        \dfrac{2}{\pi} \tan^{-1}\!\left(\dfrac{1}{\displaystyle\sum_{h=1}^{k} w_h(\mbf{u}_k)}\right) & \text{otherwise.}
    \end{cases}
\end{equation*}
The acquisition function $\alpha_\text{IWD} : \mcal{U} \rightarrow \mathbb{R}_{\geq 0}$ is defined as the weighted sum of these two complementary terms:
\begin{equation*}
    \alpha_\text{IWD}(\mbf{u}_k) = s^2(\mbf{u}_k) + \delta\, z(\mbf{u}_k),
\end{equation*}
where $\delta \geq 0$ is a hyperparameter that governs the trade-off between model-based exploitation of prediction uncertainty and geometry-based exploration of unvisited regions. 

The principal advantage of IWD$_\text{NN}$ over GS$_\text{NN}$ is its dual-objective acquisition function, which simultaneously targets regions of high model prediction error and regions of the joint space that have not been sufficiently explored, resulting in a more principled and informationally efficient data collection strategy for building thermal system identification. 
The hyperparameter $\delta$ provides an explicit and interpretable mechanism to balance exploitation and exploration, which can be tuned according to the characteristics of the building system under study. 
Nevertheless, the IWD$_\text{NN}$ weight computations scale with the number of collected samples $k$ at each active learning iteration, increasing the computational cost as the dataset grows. 
Furthermore, IWD$_\text{NN}$ requires the NN model $\hat{f}_\text{NN}$ to be sufficiently trained before the variance term $s^2(\mbf{u}_k)$ provides reliable uncertainty estimates, which makes its performance in the cold-start phase more sensitive to the quality of the initial dataset compared to the model-free GS$_\text{NN}$.

\subsection{Uncertainty Sampling}
\label{sec:uncertainty}

Uncertainty sampling selects the next control input $\mbf{u}_k^*$ by querying where the current NN model exhibits the highest predictive uncertainty. Unlike the data space methods presented in Section~\ref{sec:data_space}, uncertainty sampling is totally model-dependent, in the sense that the acquisition function is explicitly constructed from the predictions of one or more trained NN models. Two uncertainty sampling strategies are investigated in this work for the deterministic NN model: Query-by-Committee (Section~\ref{sec:qbc}) and Monte Carlo Dropout (Section~\ref{sec:mcd}).

\subsubsection{Query-by-Committee}
\label{sec:qbc}

Query-by-Committee (QBC) is an ensemble-based active learning strategy originally introduced by ~\cite{1c71439ab44149429273f1487ccd961d} and extended to regression tasks by \cite{10.5555/1777942.1777965}. 
The central principle of QBC is to maintain a committee of $N_c$ independently trained NN models and to query the candidate input that produces the highest disagreement among the committee members, under the rationale that high inter-model variance indicates a region of the input space where the current training data are insufficient to constrain the model predictions reliably. 
Each committee member $\hat{f}_{\text{NN},i}$ is trained on a bootstrap resample $\mcal{D}_i$ of the current dataset $\mcal{D}$, introducing controlled diversity among the committee members while preserving the overall distributional characteristics of the available labeled data. 

The committee of $N_c$ NN models is defined as
\begin{equation*}
    \mcal{Q}(\mbf{x}_k, \mbf{u}_k) = \left\{\hat{f}_{\text{NN},i}(\mbf{x}_k, \mbf{u}_k;\, {\theta}_i)\right\}_{i=1}^{N_c},
\end{equation*}
where ${\theta}_i$ denotes the parameter vector of the $i$-th committee member trained on the bootstrap resample $\mcal{D}_i \sim \mcal{D}$. The committee mean prediction and the inter-committee predictive variance are defined as
\begin{align*}
    \mu_\text{QBC}(\mbf{x}_k, \mbf{u}_k) &= \frac{1}{N_c} \sum_{i=1}^{N_c} \hat{f}_{\text{NN},i}(\mbf{x}_k, \mbf{u}_k;\, {\theta}_i),
    \\
    \sigma_\text{QBC}^2(\mbf{x}_k, \mbf{u}_k) &= \frac{1}{N_c} \sum_{i=1}^{N_c} \left\|\hat{f}_{\text{NN},i}(\mbf{x}_k, \mbf{u}_k;\, {\theta}_i) - \mu_\text{QBC}(\mbf{x}_k, \mbf{u}_k)\right\|_2^2,
\end{align*}
respectively. 
Therein, $\alpha_\text{QBC}(\mbf{u}_k) = \sigma_\text{QBC}^2(\mbf{x}_k, \mbf{u}_k)$.

The primary advantage of QBC is that it provides a principled, model-based measure of epistemic uncertainty derived from an ensemble of independently trained models, making it more sensitive to regions of genuine model disagreement than purely geometric approaches. However, QBC requires training $N_c$ separate NN models at each active learning iteration, resulting in a computational cost that scales linearly with the committee size $N_c$ and can become prohibitive in the context of online building thermal system identification, where computational resources and time are constrained. 
Furthermore, the quality of the uncertainty estimate depends critically on the diversity of the committee members, which in turn depends on the size and variability of the bootstrap resamples, making QBC sensitive to the size of the currently available dataset $\mcal{D}$.

\subsubsection{Monte Carlo Dropout}
\label{sec:mcd}

Monte Carlo Dropout (MCD) is an uncertainty quantification technique for deterministic neural networks proposed by \cite{pmlr-v48-gal16}, which reinterprets the standard dropout regularization mechanism as an approximate Bayesian inference procedure. 
In conventional NN training, dropout randomly deactivates a fraction of neurons during each forward pass to prevent overfitting and is typically disabled at inference time. 
In MCD, dropout is deliberately retained at inference time, so that each stochastic forward pass through the trained NN produces a different prediction by sampling a different random binary mask over the network weights. 
By performing $N_s$ such stochastic forward passes for a given input, MCD generates an approximate posterior predictive distribution over the output, from which both a mean prediction and a predictive variance can be estimated. 

Given the trained NN model $\hat{f}_\text{NN}(\cdot;{\theta})$ with dropout retained at inference time, the $i$-th stochastic forward pass produces the prediction $\hat{f}_\text{NN}(\mbf{x}_k, \mbf{u}_k;{\theta}_i)$, where ${\theta}_i$ denotes the randomly masked parameter vector at the $i$-th sample. 
The MCD predictive mean is estimated as the sample average over $N_m$ stochastic forward passes and its MCD predictive variance, which serves as the measure of epistemic uncertainty
\begin{align*}
    \mu_\text{MCD}(\mbf{x}_k, \mbf{u}_k) &= \frac{1}{N_m}\sum_{i=1}^{N_m} \hat{f}_\text{NN}(\mbf{x}_k, \mbf{u}_k;\, {\theta}_i),
    \\
    \sigma_\text{MCD}^2(\mbf{x}_k, \mbf{u}_k) &= \frac{1}{N_m}\sum_{i=1}^{N_m} \left\|\hat{f}_\text{NN}(\mbf{x}_k, \mbf{u}_k;\, {\theta}_i) - \mu_\text{MCD}(\mbf{x}_k, \mbf{u}_k)\right\|_2^2.
\end{align*}
The optimal control input is then selected by maximizing the MCD predictive variance over the admissible control input space
\begin{equation*}
    \mbf{u}_k^* = \underset{\mbf{u}_k \in \mcal{U}}{\arg\max}\; \sigma_\text{MCD}^2(\mbf{x}_k, \mbf{u}_k), \; \; \text{with} \; \; \alpha_\text{MCD}(\mbf{u}_k) = \sigma_\text{MCD}^2(\mbf{x}_k, \mbf{u}_k).
\end{equation*}

The key advantage of MCD over QBC is its ability to produce calibrated uncertainty estimates from a single trained NN model, thereby avoiding the computational overhead of training and maintaining multiple committee members. This makes MCD particularly well-suited to online active learning scenarios in building energy system identification, where model retraining must occur frequently as new observations are collected. 
Nevertheless, MCD introduces its own limitations: the quality of the uncertainty estimates is sensitive to the dropout rate hyperparameter, which must be carefully tuned, and the theoretical connection between MC dropout and Bayesian inference holds only approximately, meaning that the resulting variance estimates may underestimate the true predictive uncertainty, particularly in regions of the input space that are far from the training distribution~\cite{pmlr-v48-gal16}.

\subsection{Expected Model Change}
\label{sec:model_change}

The expected model change category selects the next control input $\mbf{u}_k^*$ by identifying the candidate that is anticipated to induce the largest update to the current NN model parameters, under the rationale that samples which cause the greatest parameter change are the most informative for reducing the generalization error of the model. 
Two methods in this category are investigated: the Expected Model Change Maximization method (Section~\ref{sec:EMCM}) and the Maximize Last Layer Change method (Section~\ref{sec:mllc}).

\subsubsection{Expected Model Change Maximization}
\label{sec:EMCM}

The Expected Model Change Maximization (EMCM) method was originally proposed by \cite{Cai2013MEMC} as an active learning framework for regression that selects the training sample anticipated to induce the largest change to the current model parameters. 
The core motivation of EMCM is rooted in the Stochastic Gradient Descent (SGD) update rule: a candidate sample that produces a large gradient of the loss function with respect to the model parameters will cause a large parameter update, and is therefore deemed maximally informative for improving the model.
Since the true label of a candidate sample is unknown prior to querying, a bootstrap ensemble $\mcal{B}(K) = \{\hat{f}_{\text{NN},1}, \dots, \hat{f}_{\text{NN},K}\}$ of $K$ NN models, each trained on a bootstrap resample of the current dataset $\mcal{D}$, is used to generate a distribution of hypothetical labels, from which the expected model change is approximated. 

Given the current NN model $\hat{f}_\text{NN}(\cdot\,;{\theta}_k)$ parameterized by ${\theta}_k \in \mathbb{R}^{n_\theta}$, consider a candidate sample $[\mbf{x}_k, \mbf{u}_k]$ with hypothetical successor state $\mbf{x}_{k+1}^+$. If this candidate were added to the training dataset, the resulting mean squared error loss on the augmented dataset $\mcal{D}^+ = \mcal{D} \cup \{[\mbf{x}_k, \mbf{u}_k]; \mbf{x}_{k+1}^+\}$ would include the additional term
\begin{equation*}
    \ell\!\left({\theta}_k, \mbf{x}_{k+1}^+\right) = \left\|\mbf{x}_{k+1}^+ - \hat{f}_\text{NN}(\mbf{x}_k, \mbf{u}_k;\, {\theta}_k)\right\|_2^2, 
\end{equation*}
The model change $\mcal{C}(\mbf{u}_k, \mbf{x}_{k+1}^+)$ induced by incorporating this candidate is estimated via the SGD gradient of this loss with respect to ${\theta}_k$
\begin{equation*}
    \mcal{C}\!\left(\mbf{u}_k, \mbf{x}_{k+1}^+\right) = \left\|\frac{\partial\, \ell\!\left({\theta}_k, \mbf{x}_{k+1}^+\right)}{\partial\, {\theta}_k}\right\|_2.
\end{equation*}
Since $\mbf{x}_{k+1}^+$ is unknown before querying, the expected model change is approximated by averaging over the $K$ hypothetical labels $\{\hat{y}_k^{(j)}\}_{j=1}^K$ generated by the bootstrap ensemble $\mcal{B}(K)$, where $\hat{y}_k^{(j)} = \hat{f}_{\text{NN},j}(\mbf{x}_k, \mbf{u}_k;\, {\theta}_k^{(j)})$
\begin{equation*}
    \alpha_\text{EMCM}(\mbf{u}_k) =  \mathbb{E}\!\left[\mcal{C}(\mbf{u}_k)\right] = \frac{1}{K}\sum_{j=1}^{K} \left\|\frac{\partial\, \ell\!\left({\theta}_k,\, \hat{y}_k^{(j)}\right)}{\partial\, {\theta}_k}\right\|_2.
\end{equation*}

EMCM offers a theoretically principled connection between sample informativeness and model parameter change, providing a stronger justification for sample selection than purely geometric approaches.
However, EMCM requires constructing and evaluating a bootstrap ensemble of $K$ NN models at each active learning iteration, resulting in a computational overhead that scales linearly with $K$. 
Furthermore, computing the full gradient $\partial\,\ell / \partial\,{\theta}_k$ with respect to all NN parameters is considerably more expensive than computing only the last-layer gradient, which motivates the MLLC method described in the following subsection. 
Additionally, the gradient-based model change estimate is a local linear approximation of the true parameter shift, and may become unreliable in highly nonlinear regions of the loss landscape that are characteristic of deep NN models used for building thermal system identification.

\subsubsection{Maximize Last Layer Change}
\label{sec:mllc}

The Maximize Last Layer Change (MLLC) method is derived from the deep batch active learning framework of \cite{Ash2020Deep}, which selects batches using diverse, uncertain gradient embeddings; MLLC adapts this framework to the regression setting by retaining solely the last-layer gradient norm as the acquisition criterion and discarding the batch diversity sampling step.
The theoretical motivation for MLLC is grounded in Proposition 1 of \cite{Ash2020Deep}, which establishes that the gradient norm of the loss with respect to the weights of the final output layer, computed using the model's current predicted label as a proxy for the unknown true label, provides a lower bound on the gradient norm induced by any other possible label. 
This implies that a large last-layer gradient norm is a conservative but valid indicator of high model uncertainty: regions of the input space where the NN is uncertain produce large predicted residuals and therefore large last-layer gradients, while regions where the model is confident produce near-zero residuals and small gradients.
By restricting the gradient computation to the last layer only, MLLC achieves a substantially lower computational cost per active learning iteration compared to EMCM, without sacrificing the connection between gradient magnitude and model informativeness.
Let the NN $\hat{f}_\text{NN}(\cdot\,;{\theta}_k)$ be decomposed as $\hat{f}_\text{NN}(\mbf{x}_k, \mbf{u}_k;{\theta}_k) = \mbf{W}_\text{out}\,\boldsymbol{\phi}(\mbf{x}_k, \mbf{u}_k;{\theta}_\text{hidden}) + \boldsymbol{b}_\text{out}$, where $\boldsymbol{\phi}(\cdot\,;{\theta}_\text{hidden}) \in \mathbb{R}^{n_h}$ denotes the output of the penultimate (second-to-last) layer, referred to as the feature embedding, and $\mbf{W}_\text{out} \in \mathbb{R}^{n_x \times n_h}$ denotes the weight matrix and $\mbf{b}_\text{out} \in \mathbb{R}^{n_x}$ denotes the bias vector of the final output layer. 

Utilizing the bootstrap technique from EMCM, the MLLC acquisition function is then the L2-norm of this last-layer gradient embedding, which serves as a computationally efficient proxy for overall model uncertainty at the candidate input
\begin{equation*}
    \alpha_\text{MLLC}(\mbf{u}_k) = \frac{1}{K}\sum_{j=1}^{K} \left\|\frac{\partial\, \ell\!\left({\theta_\text{out}}_k,\, \hat{y}_k^{(j)}\right)}{\partial\, {\theta_\text{out}}_k}\right\|_2, \quad \theta_\text{out} = [\mbf{W}_\text{out}, \mbf{b}_\text{out}],
\end{equation*}
where $\theta_{\text{out}_k}$ represents the last layer parameters at the time step $k$.
The key advantage of MLLC over EMCM is its substantially reduced computational cost, since only the gradient with respect to the final layer weights and biases is required, rather than the full gradient with respect to all $n_\theta$ parameters of the NN. This makes MLLC significantly more scalable to deep NN architectures with large numbers of hidden layers and parameters, which are commonly employed in building energy system identification tasks. 
However, restricting the gradient to the last layer alone may underestimate the total model change induced by a candidate sample in deep networks.

%% file: stochastic.tex
\section{Active Learning for Stochastic Models}
\label{sec:stochastic_al}

This section presents the active learning methods tailored to the GP model.
Unlike the deterministic NN, the GP natively provides a posterior predictive distribution at every test point, yielding both a mean prediction $\mu_\text{GP}$ and a predictive variance $\sigma_\text{GP}^2$ in closed form (Section~\ref{sec:gp}).
The active learning acquisition functions for the stochastic model directly exploit this probabilistic structure, without requiring ensemble approximations or stochastic inference mechanisms.
The GP-based methods are organized into four categories: data space (Section~\ref{sec:data_space_gp}), uncertainty sampling (Section~\ref{sec:uncertainty_gp}), information gain (Section~\ref{sec:info_gain_gp}), and model change (Section~\ref{sec:model_change_gp}).

\subsection{Data Space}
\label{sec:data_space_gp}

Two data space strategies are investigated for the GP model: Greedy Sampling (Section~\ref{sec:gs_gp}) and Inverse Distance Weighting (Section~\ref{sec:idw_gp}), both of which adapt their NN-based counterparts from Section~\ref{sec:data_space} by substituting the GP posterior mean for the NN point prediction wherever a model-based term is required.

\subsubsection{Greedy Sampling}
\label{sec:gs_gp}

The greedy sampling method for the GP model, denoted GS$_\text{GP}$, operates identically to GS$_\text{NN}$ presented in Section~\ref{sec:greedy}.
Since the acquisition function of GS relies solely on the geometric structure of the joint state-input space $\mcal{Z}$ and does not involve any model predictions, it is fully model-agnostic and requires no adaptation for the GP framework.
The acquisition function and optimal control input selection therefore follow the same formulation as in Section~\ref{sec:greedy},
with the subscript GP used solely to distinguish GS$_\text{GP}$ from the NN counterpart in the experimental comparison.
The advantages and limitations of GS$_\text{GP}$ are identical to those of GS$_\text{NN}$ as analyzed in Section~\ref{sec:greedy}.

\subsubsection{Inverse Distance Weighting}
\label{sec:idw_gp}

The Inverse Distance Weighting method adapted for the GP model, denoted IWD$_\text{GP}$, follows the same dual-objective acquisition structure as IWD$_\text{NN}$ presented in Section~\ref{sec:idw}.
The sole modification consists of replacing the NN mean prediction $\hat{f}_\text{NN}$ in the IDW variance term \eqref{eq:IDW-NN-S} with the GP posterior mean $\mu_\text{GP}$ \eqref{eq:gp_mean}.
All remaining components of the acquisition function, namely the IDW weight function $w_j$, the normalized IDW coefficient $v_j$, and the exploration function $z$, are defined identically as in Section~\ref{sec:idw} and are not restated here.

The IDW variance function for the GP model is defined as
\begin{equation*}
    s_\text{GP}^2(\mbf{u}_k) = \sum_{j=1}^{k} v_j(\mbf{u}_k)\, \big\|\mbf{x}_{j} - \mu_\text{GP}\!\left(\mbf{z}_{j-1};\, \boldsymbol{\theta}_k\right)\big\|_2^2,
\end{equation*}
where $\mu_\text{GP}(\mbf{z}_{j-1};\, \boldsymbol{\theta}_k)$ is the GP posterior mean prediction at the input $\mbf{z}_{j-1} = [\mbf{x}_{j-1}, \mbf{u}_{j-1}]$ under the hyperparameters $\boldsymbol{\theta}_k$ fitted on $\mcal{D}_k$.
The acquisition function and optimal control input are then
\begin{align*}
    \alpha_\text{IWD}(\mbf{u}_k) &= s_\text{GP}^2(\mbf{u}_k) + \delta\, z(\mbf{u}_k),
\end{align*}
where $\delta \geq 0$ governs the trade-off between exploitation of GP prediction error and geometric exploration of unvisited regions.



\subsection{Uncertainty Sampling}
\label{sec:uncertainty_gp}

Uncertainty sampling for the GP model directly exploits the posterior predictive variance $\sigma_\text{GP}^2$ as the primary measure of model uncertainty.
Since the GP yields $\sigma_\text{GP}^2$ analytically in closed form at every test point, no ensemble approximations or stochastic inference mechanisms are required, in contrast to QBC and MCD for the NN model (Section~\ref{sec:uncertainty}).
Two uncertainty sampling strategies are investigated in this work for the GP model: Maximize Variance (Section~\ref{sec:mv}) and Integrated Variance Reduction (Section~\ref{sec:ivr}).

\subsubsection{Maximize Variance}
\label{sec:mv}

The Maximize Variance (MV) method, investigated by~\cite{pmlr-v120-buisson-fenet20a} in the context of actively learning GP dynamics, is the most direct application of the GP's native uncertainty to active learning.
The core principle is to select the control input $\mbf{u}_k$ that maximizes the GP posterior predictive variance at the current system state $\mbf{x}_k$, under the rationale that regions of high variance correspond to parts of the input space where the current training data $\mcal{D}_k$ provide the least constraint on the GP posterior.

The MV acquisition function is defined directly from the GP posterior predictive variance \eqref{eq:gp_var}
\begin{equation*}
    \alpha_\text{MV}(\mbf{u}_k) = \sigma_\text{GP}^2(\mbf{x}_k, \mbf{u}_k;\, \boldsymbol{\theta}_k).
\end{equation*}

The principal advantage of MV is its simplicity and computational efficiency: it requires no additional computation beyond the GP posterior inference already performed at each active learning step.
However, MV is a purely local, myopic strategy that does not account for the global impact of a new observation on the GP posterior variance across the entire input space.
Selecting the point of highest current local variance does not necessarily maximize the total reduction in posterior uncertainty, since the variance reduction induced by a new observation propagates to neighboring regions of the input space through the kernel structure.
A different query point may therefore produce a larger integrated variance reduction across $\mcal{U}$, which motivates the IVR method described in the following section.

\subsubsection{Integrated Variance Reduction}
\label{sec:ivr}

The Integrated Variance Reduction (IVR) method, proposed by~\cite{kontoudis2023closed} under the name Active Learning Cohn (ALC) with GP surrogates, addresses the myopic limitation of MV by selecting the control input $\mbf{u}_k$ that maximizes the total reduction in GP posterior predictive variance integrated over the full admissible control input space $\mcal{U}$.
Rather than querying the point of highest local variance, IVR accounts for the global effect of a new observation on the GP posterior, thereby promoting a more globally efficient experimental design.

The GP posterior variance at a test point $\mbf{z}' = [\mbf{x}_k, \mbf{u}']$ is updated when a new noisily observed sample at $\mbf{z}_\text{new} = [\mbf{x}_k, \mbf{u}_k]$ is added to the training dataset $\mcal{D}_k$ according to the rank-one update formula
\begin{equation}
    \sigma^2\!\left(\mbf{z}';\, \mbf{z}_\text{new},\, \boldsymbol{\theta}_k\right) = \sigma_\text{GP}^2\!\left(\mbf{z}';\, \boldsymbol{\theta}_k\right) - \frac{c_k\!\left(\mbf{z}', \mbf{z}_\text{new}\right)^2}{\sigma_\text{GP}^2\!\left(\mbf{z}_\text{new};\, \boldsymbol{\theta}_k\right) + \sigma_n^2}, \label{eq:gp_var_update}
\end{equation}
where $c_k(\mbf{z}', \mbf{z}_\text{new})$ is the GP posterior covariance between $\mbf{z}'$ and $\mbf{z}_\text{new}$ given the current dataset $\mcal{D}_k$, defined as
\begin{equation}
    c_k\!\left(\mbf{z}', \mbf{z}_\text{new}\right) = k\!\left(\mbf{z}', \mbf{z}_\text{new}\right) - \mbf{k}_*'^\top\, \mbf{K}_{y,k}^{-1}\, \mbf{k}_\text{new}, \label{eq:post_cov}
\end{equation}
where $\mbf{k}_*' = [k(\mbf{z}', \mbf{z}_0), \ldots, k(\mbf{z}', \mbf{z}_{k-1})]^\top \in \mathbb{R}^k$; $\mbf{k}_\text{new} = [k(\mbf{z}_\text{new}, \mbf{z}_0), \ldots, k(\mbf{z}_\text{new},  \mbf{z}_{k-1})]^\top \in \mathbb{R}^k$ are the cross-covariance vectors of $\mbf{z}'$ and $\mbf{z}_\text{new}$ with the training inputs $\mbf{Z}_k$, respectively.
The denominator $\sigma_\text{GP}^2(\mbf{z}_\text{new};\, \boldsymbol{\theta}_k) + \sigma_n^2$ in \eqref{eq:gp_var_update} represents the total predictive variance at the query point, including the observation noise $\sigma_n^2$.

The IVR acquisition function is defined as the integral of the expected variance reduction \eqref{eq:gp_var_update} over the admissible control input space
\begin{align}
    \alpha_\text{IVR}(\mbf{u}_k) &= \int_{\mcal{U}} \left[\sigma_\text{GP}^2(\mbf{x}_k, \mbf{u}';\, \boldsymbol{\theta}_k) - \sigma^2\!\left(\mbf{x}_k, \mbf{u}';\, \mbf{z}_\text{new},\, \boldsymbol{\theta}_k\right)\right] d\mbf{u}' \notag \\
    &= \int_{\mcal{U}} \frac{c_k\!\left([\mbf{x}_k, \mbf{u}'],\, \mbf{z}_\text{new}\right)^2}{\sigma_\text{GP}^2\!\left(\mbf{z}_\text{new};\, \boldsymbol{\theta}_k\right) + \sigma_n^2}\; d\mbf{u}'. \label{eq:ivr_main}
\end{align}
Since the denominator in \eqref{eq:ivr_main} does not depend on the integration variable $\mbf{u}'$, the acquisition function simplifies to
\begin{equation}
    \alpha_\text{IVR}(\mbf{u}_k) = \frac{1}{\sigma_\text{GP}^2\!\left(\mbf{z}_\text{new};\, \boldsymbol{\theta}_k\right) + \sigma_n^2} \int_{\mcal{U}} c_k\!\left([\mbf{x}_k, \mbf{u}'],\, \mbf{z}_\text{new}\right)^2 d\mbf{u}'. \label{eq:ivr_simplified}
\end{equation}
For the RBF kernel \eqref{eq:rbf}, the posterior covariance $c_k([\mbf{x}_k, \mbf{u}'], \mbf{z}_\text{new})$ is a smooth, squared-exponential function of $\mbf{u}'$, and the integral in \eqref{eq:ivr_simplified} admits a closed-form expression when $\mcal{U}$ is an axis-aligned box.
Following~\cite{kontoudis2023closed}, the closed-form gradient $\partial\, \alpha_\text{IVR} / \partial\, \mbf{u}_k$ is derived analytically for the separable RBF kernel, enabling efficient gradient-based maximization of the IVR acquisition function without resorting to computationally expensive grid evaluations over $\mcal{U}$.

The key advantage of IVR over MV is its global perspective: rather than myopically selecting the point of highest current local variance, IVR accounts for the variance-reducing effect of the new observation across the entire admissible space $\mcal{U}$, thereby identifying a more globally informative query point.
The closed-form gradient derived in~\cite{kontoudis2023closed} makes optimization of $\alpha_\text{IVR}$ computationally tractable compared to a naive numerical integration approach.
Nevertheless, evaluating $\alpha_\text{IVR}$ and its gradient requires computing the posterior covariance $c_k$ in \eqref{eq:post_cov}, which involves the inverse of the $k \times k$ noisy kernel matrix $\mbf{K}_{y,k}$.
Furthermore, the closed-form integral derivation is specific to the RBF kernel, and extending IVR to other kernel families requires re-deriving the corresponding analytical expressions from first principles.

\subsection{Information Gain}
\label{sec:info_gain_gp}

Two information gain criteria are investigated for the GP model: Fisher Information (Section~\ref{sec:fi}) and Shannon Entropy (Section~\ref{sec:se}), both of which select experiments based on their expected contribution to estimating the GP kernel hyperparameters, rather than the predictive uncertainty at a single candidate point.

\subsubsection{Fisher Information}
\label{sec:fi}

Rather than targeting regions of high predictive uncertainty, Fisher Information (FI) selects the candidate input $\mbf{u}_k$ that maximizes the determinant (for D-optimal) of the Fisher information matrix (FIM) of the GP log marginal likelihood with respect to the kernel hyperparameters, thereby identifying the experiment most informative for estimating the GP model structure from the available data.

Consider the dataset augmented with a candidate observation at $\mbf{z}_\text{new} = [\mbf{x}_k, \mbf{u}_k]$, denoted $\mcal{D}_k^+ = \mcal{D}_k \cup \{\mbf{z}_\text{new}\}$, and let $\mbf{K}_{y,k}^+ = \mbf{K}_k^+ + \sigma_n^2\, \mbf{I}_{k+1} \in \mathbb{R}^{(k+1)\times(k+1)}$ denote the corresponding noisy kernel matrix.
The kernel hyperparameters are expressed in log-space as $\boldsymbol{\theta} = \{\log\sigma_f^2,\, \log\ell_1, \ldots, \log\ell_{n_z},\, \log\sigma_n^2\} \in \mathbb{R}^p$, where $p = n_z + 2$, to ensure positivity and improve numerical conditioning.
The FIM $\mbf{F}(\mcal{D}_k^+) \in \mathbb{R}^{p \times p}$ is defined entrywise as \cite{LY201740}
\begin{equation}
    [\mbf{F}(\mcal{D}_k^+)]_{ij} = \frac{1}{2}\,\mathrm{tr}\!\left(\mbf{K}_{y,k}^{+,-1}\frac{\partial \mbf{K}_{y,k}^+}{\partial \theta_i}\,\mbf{K}_{y,k}^{+,-1}\frac{\partial \mbf{K}_{y,k}^+}{\partial \theta_j}\right), \label{eq:fim}
\end{equation}
where $\partial \mbf{K}_{y,k}^+ / \partial \theta_i$ denotes the matrix of partial derivatives of $\mbf{K}_{y,k}^+$ with respect to the $i$-th log-space hyperparameter.
For the ARD-RBF kernel \eqref{eq:rbf}, these partial derivatives are
\begin{align*}
    \frac{\partial \mbf{K}_{y,k}^+}{\partial \log\sigma_f^2} &= \mbf{K}_f^+, \\
    \frac{\partial \mbf{K}_{y,k}^+}{\partial \log\ell_l} &= \mbf{K}_f^+ \odot \frac{\mbf{R}_l^+}{\ell_l^2}, \quad l = 1, \ldots, n_z, \\
    \frac{\partial \mbf{K}_{y,k}^+}{\partial \log\sigma_n^2} &= \sigma_n^2\, \mbf{I}_{k+1},
\end{align*}
where $[\mbf{K}_f^+]_{ij} = \sigma_f^2 \exp\!\left(-\|\mbf{z}_i - \mbf{z}_j\|_2^2 / 2\ell^2\right)$ is the noise-free kernel matrix of $\mcal{D}_k^+$, $[\mbf{R}_l^+]_{ij} = (z_i^{(l)} - z_j^{(l)})^2$ is the squared distance matrix along the $l$-th input dimension, and $\odot$ denotes the Hadamard (element-wise) product.

The FI acquisition function is the D-optimality criterion evaluated on the augmented dataset
\begin{equation*}
    \alpha_\text{FI}(\mbf{u}_k) = \det\!\left(\mbf{F}(\mcal{D}_k^+)\right).
\end{equation*}

The D-optimality criterion has a direct geometric interpretation: maximizing $\det(\mbf{F})$ is equivalent to maximizing the volume of the confidence ellipsoid in hyperparameter space, thereby selecting the experiment that most tightly constrains the estimate of $\boldsymbol{\theta}$.
Other classical criteria from optimal experimental design, such as A-optimality ($\mathrm{tr}(\mbf{F})$) and E-optimality 
($\lambda_{\min}(\mbf{F})$), can be substituted as the acquisition criterion with analogous interpretations.

The primary advantage of FI is its principled grounding in hyperparameter estimation theory: by directly maximizing the Fisher information about $\boldsymbol{\theta}$, FI selects experiments that improve the quality of the GP model structure, rather than simply targeting regions of high predictive uncertainty as in MV or IVR.
This makes FI particularly well-suited to the early phases of the active learning process, where hyperparameter uncertainty is large and the GP model structure is poorly identified.
Nevertheless, the FIM quantifies only the information about hyperparameters $\boldsymbol{\theta}$, and does not directly account for predictive uncertainty at the candidate point, which can lead to suboptimal sample allocation in regions where the GP model is already well-specified but the predictive variance remains high.

\subsubsection{Shannon Entropy}
\label{sec:se}

The Shannon Entropy (SE) method, originally proposed by~\cite{8443729} for GP-based optimal experiment design, selects the control input $\mbf{u}_k$ that maximizes the ratio of the marginal predictive variance $\tilde{\sigma}^2(\mbf{z}_*)$ to the standard GP predictive variance $\sigma_\text{GP}^2(\mbf{z}_*)$.
The marginal predictive variance accounts for the additional uncertainty in the GP prediction arising from imprecise knowledge of the kernel hyperparameters $\boldsymbol{\theta}$, quantified through a Laplace approximation of the hyperparameter posterior.
This ratio captures both uncertainty from data scarcity, encoded in $\sigma_\text{GP}^2$, and uncertainty from hyperparameter ignorance, encoded in the additional terms of $\tilde{\sigma}^2$, making SE fundamentally distinct from MV which captures only the former.

To quantify hyperparameter uncertainty, the posterior $p(\boldsymbol{\theta} \mid \mcal{D}_k)$ is approximated by a Gaussian centered at the MAP estimate $\boldsymbol{\theta}_k$ with covariance $\boldsymbol{\Sigma}_k$.
The MAP estimate is obtained by maximizing the log marginal likelihood with a Gaussian prior of precision $\lambda > 0$ on $\boldsymbol{\theta}$:
\begin{equation*}
    \boldsymbol{\theta}_k = \underset{\boldsymbol{\theta}}{\arg\max}\left[\log p(\mbf{y}_k \mid \mbf{Z}_k, \boldsymbol{\theta}) - \frac{\lambda}{2}\|\boldsymbol{\theta}\|_2^2\right].
\end{equation*}
The Laplace approximation at $\boldsymbol{\theta}_k$ gives the posterior covariance
\begin{equation}
    \boldsymbol{\Sigma}_k = \left(\mbf{H}_k + \lambda\, \mbf{I}_p\right)^{-1}, \label{eq:laplace}
\end{equation}
where $\mbf{H}_k = -\partial^2 \log p(\mbf{y}_k \mid \mbf{Z}_k, \boldsymbol{\theta}) / \partial \boldsymbol{\theta}^2\big|_{\boldsymbol{\theta} = \boldsymbol{\theta}_k} \in \mathbb{R}^{p \times p}$ is the Hessian of the negative log marginal likelihood evaluated at the MAP estimate, and the regularization term $\lambda \mbf{I}_p$ ensures positive definiteness of $\mbf{H}_k + \lambda\mbf{I}_p$ when the dataset is small and the Hessian is poorly conditioned.

Given the Laplace posterior $\boldsymbol{\Sigma}_k$, the marginal predictive variance at $\mbf{z}_* = [\mbf{x}_k, \mbf{u}_k]$, which integrates out the uncertainty in $\boldsymbol{\theta}$, is approximated by~\cite{8443729} as
\begin{equation}
    \tilde{\sigma}^2(\mbf{z}_*;\, \boldsymbol{\theta}_k) = \frac{4}{3}\,\sigma_\text{GP}^2(\mbf{z}_*;\, \boldsymbol{\theta}_k) + \mbf{g}_\mu^\top \boldsymbol{\Sigma}_k\, \mbf{g}_\mu + \frac{\mbf{g}_{\sigma^2}^\top \boldsymbol{\Sigma}_k\, \mbf{g}_{\sigma^2}}{3\,\sigma_\text{GP}^2(\mbf{z}_*;\, \boldsymbol{\theta}_k)}, \label{eq:mgp_var}
\end{equation}
where $\mbf{g}_\mu = \partial \mu_\text{GP}(\mbf{z}_*;\, \boldsymbol{\theta}) / \partial \boldsymbol{\theta}\big|_{\boldsymbol{\theta}_k} \in \mathbb{R}^p$ and $\mbf{g}_{\sigma^2} = \partial \sigma_\text{GP}^2(\mbf{z}_*;\, \boldsymbol{\theta}) / \partial \boldsymbol{\theta}\big|_{\boldsymbol{\theta}_k} \in \mathbb{R}^p$ are the gradients of the GP posterior mean and variance with respect to the hyperparameters, evaluated at the MAP estimate.
The three terms in \eqref{eq:mgp_var} have distinct physical interpretations: $\frac{4}{3}\sigma_\text{GP}^2$ is the base GP predictive variance scaled by a factor of $4/3$, which accounts for the inherent variance of the variance estimator itself; 
$\mbf{g}_\mu^\top \boldsymbol{\Sigma}_k\, \mbf{g}_\mu$ is the additional variance in the GP predictive mean induced by uncertainty in $\boldsymbol{\theta}$, propagated via the delta method: candidate points where the mean prediction is highly sensitive to hyperparameter changes contribute large values to this term;
$(\mbf{g}_{\sigma^2}^\top \boldsymbol{\Sigma}_k\, \mbf{g}_{\sigma^2}) / (3\sigma_\text{GP}^2)$ captures the additional uncertainty arising from the sensitivity of the predictive variance $\sigma_\text{GP}^2$ itself to hyperparameter changes, normalized by the current variance level.

The SE acquisition function is defined as the ratio of the marginal predictive variance to the standard GP predictive variance
\begin{equation*}
    \alpha_\text{SE}(\mbf{u}_k) = \frac{\tilde{\sigma}^2(\mbf{x}_k, \mbf{u}_k;\, \boldsymbol{\theta}_k)}{\sigma_\text{GP}^2(\mbf{x}_k, \mbf{u}_k;\, \boldsymbol{\theta}_k)}.
\end{equation*}

The ratio $\tilde{\sigma}^2 / \sigma_\text{GP}^2$ measures the total predictive entropy relative to the purely aleatoric entropy, and exceeds unity whenever hyperparameter uncertainty contributes additional variance beyond the standard GP posterior.
The principal advantage of SE over MV and IVR is its explicit quantification of hyperparameter epistemic uncertainty: by accounting for $\boldsymbol{\Sigma}_k$ through the delta method, SE selects experiments that reduce both predictive and structural uncertainty of the GP model simultaneously.
However, computing $\boldsymbol{\Sigma}_k = (\mbf{H}_k + \lambda \mbf{I}_p)^{-1}$ requires second-order differentiation of the log marginal likelihood at each active learning step, resulting in a computational cost that scales as $\mcal{O}(p^2 k^3)$ in total, substantially higher than that of MV.
Furthermore, the Laplace approximation underlying \eqref{eq:laplace} is accurate only when the log marginal likelihood is well-approximated by a quadratic function near $\boldsymbol{\theta}_k$, which may not hold when the posterior over $\boldsymbol{\theta}$ is multimodal or highly skewed, as can occur with limited training data in the early stages of online building energy system identification.

\subsection{Model Change}
\label{sec:model_change_gp}

The model change category for the GP model selects the control input $\mathbf{u}_k$ that is anticipated to induce the largest update to the GP posterior, in direct analogy with the EMCM and MLLC methods for the NN model (Section~\ref{sec:model_change}).
Unlike those gradient-based methods, however, the GP's analytical posterior provides closed-form acquisition functions that quantify model change through the lens of Bayesian optimization.
Two strategies are investigated: the Upper Confidence Bound (Section~\ref{sec:ucb}) and Probability of Improvement (Section~\ref{sec:pi}), both of which balance exploitation of the current GP mean prediction with exploration of regions of high posterior uncertainty.

\subsubsection{Upper Confidence Bound}
\label{sec:ucb}

The Upper Confidence Bound (UCB) acquisition function originates from the GP-UCB algorithm for Bayesian optimization \cite{wang2023recent} and has been extended to active learning of conditional mean embeddings by~\cite{pmlr-v124-ray-chowdhury20a}.
In the context of active learning for building thermal system identification, UCB selects the control input that jointly maximizes the GP posterior mean prediction and the posterior standard deviation, thereby simultaneously targeting regions of high predicted system response (exploitation) and regions of high model uncertainty where the GP posterior is most likely to change upon receiving a new observation (exploration).

The UCB acquisition function is defined as
\begin{equation}
    \alpha_\text{UCB}(\mathbf{u}_k) = \mu_\text{GP}(\mathbf{x}_k, \mathbf{u}_k;\, \boldsymbol{\theta}_k) + \beta\, \sigma_\text{GP}(\mathbf{x}_k, \mathbf{u}_k;\, \boldsymbol{\theta}_k), \label{eq:ucb}
\end{equation}
where $\mu_\text{GP}(\mathbf{z}_*;\, \boldsymbol{\theta}_k)$ is the GP posterior predictive mean \eqref{eq:gp_mean}, $\sigma_\text{GP}(\mathbf{z}_*;\, \boldsymbol{\theta}_k) = \sqrt{\sigma_\text{GP}^2(\mathbf{z}_*;\, \boldsymbol{\theta}_k)}$ is the posterior predictive standard deviation, and $\beta \geq 0$ is a hyperparameter governing the exploration-exploitation trade-off.

The role of $\beta$ in \eqref{eq:ucb} provides an explicit and interpretable mechanism for controlling the sampling behavior: when $\beta = 0$, UCB reduces to pure exploitation by greedily selecting the input with the highest predicted successor state, regardless of uncertainty; as $\beta$ increases, the exploration term $\beta\,\sigma_\text{GP}$ becomes dominant and UCB increasingly prioritizes unvisited regions of $\mathcal{U}$, approaching the behavior of MV in the limit of large $\beta$.
The connection to model change arises through the standard deviation term $\sigma_\text{GP}$: selecting inputs with high posterior uncertainty directly targets regions where the GP posterior will undergo the most significant update upon observing the system response, thereby driving efficient model improvement.

The primary advantage of UCB is its computational simplicity: both $\mu_\text{GP}$ and $\sigma_\text{GP}$ are computed directly from the GP posterior inference already performed at each active learning step, with no additional quantities required.
The acquisition function is smooth and differentiable with respect to $\mathbf{u}_k$ for the RBF kernel, enabling efficient gradient-based maximization.
Furthermore, the $\beta$ hyperparameter provides a principled and application-specific mechanism for tuning the degree of exploration, which can be adjusted according to the operational constraints and data budget available in the building energy identification task.
However, UCB's exploitation term $\mu_\text{GP}$ introduces a dependence on the absolute magnitude of the predicted system response, which can lead to biased sampling in building systems with heterogeneous state dynamics across the state-input space.
Specifically, UCB may persistently oversample high-response regions even when those regions are already well-characterized by the current GP model, because the exploitation term continues to reward them even after the uncertainty $\sigma_\text{GP}$ has been reduced.
The choice of $\beta$ also requires careful tuning: an undersized $\beta$ leads to premature exploitation and poor model coverage, while an oversized $\beta$ produces behavior indistinguishable from MV without benefiting from the mean prediction signal.

\subsubsection{Probability of Improvement}
\label{sec:pi}

The Probability of Improvement (PI) acquisition function originates from the Bayesian optimization literature \cite{frazier2018tutorialbayesianoptimization} as a method for sequentially selecting experiments to maximize an unknown objective function, and is adopted here for active learning of building thermal dynamics.
In the present context, PI selects the control input $\mathbf{u}_k$ that maximizes the posterior probability that the predicted system response exceeds the best output value observed thus far, directly targeting candidate inputs most likely to produce an improvement over the current best observation.

Let $y_k^* = \max_{i < k} y_i$ denote the best output value observed in the training dataset $\mathcal{D}_k$ up to step $k$, and define the standardized improvement ratio
\begin{equation*}
    Z_k(\mathbf{u}_k) = \frac{\mu_\text{GP}(\mathbf{x}_k, \mathbf{u}_k;\, \boldsymbol{\theta}_k) - y_k^* - \xi}{\sigma_\text{GP}(\mathbf{x}_k, \mathbf{u}_k;\, \boldsymbol{\theta}_k)},
\end{equation*}
where $\xi \geq 0$ is an exploration parameter that shifts the improvement threshold above the current best observation, encouraging the algorithm to seek points that exceed $y_k^*$ by a margin of at least $\xi$.
The PI acquisition function is then given in closed form by \cite{danka2018modalmodularactivelearning}
\begin{equation}
    \alpha_\text{PI}(\mathbf{u}_k) = \Phi\!\left(Z_k(\mathbf{u}_k)\right), \label{eq:pi}
\end{equation}
where $\Phi(\cdot)$ denotes the standard Gaussian cumulative distribution function.

The acquisition function in \eqref{eq:pi} reports the posterior probability that the candidate input produces a successor state exceeding the improvement threshold $y_k^* + \xi$, computed directly from the standardized distance $Z_k$ between the predicted mean and that threshold.
This probability grows large whenever $\mu_\text{GP} \gg y_k^*$ and the prediction is confident, since $Z_k$ becomes large and positive and $\Phi(Z_k) \rightarrow 1$, but it also grows when $\sigma_\text{GP}$ is large even though $\mu_\text{GP}$ remains close to $y_k^*$, because a wide predictive distribution assigns substantial probability mass beyond the threshold regardless of its mean.
When $\xi = 0$, \eqref{eq:pi} reduces to the standard noise-free PI formula of \cite{pmlr-v222-zhou24a}.

The principal advantage of PI over UCB is its direct probabilistic interpretation: rather than requiring manual tuning of a single $\beta$ weight, the PI formula reports the exact posterior probability of improvement through the Gaussian CDF, yielding an acquisition function that is straightforward to compute and interpret for practitioners.
Furthermore, the exploration parameter $\xi$ offers an explicit and tunable mechanism for shifting the improvement threshold, allowing the algorithm to seek either marginal or substantial gains over $y_k^*$ depending on the building system under study.
Nevertheless, PI inherits the limitation that $y_k^*$ is defined relative to a single scalar summary of the training outputs, which may not fully characterize the quality of the current GP model for all regions of the state-input space.
In the multi-output setting of this work, where $n_x$ independent GPs are trained simultaneously, $y_k^*$ is computed per output dimension, and the aggregated PI across dimensions requires a representative single GP as described in Section~\ref{sec:stochastic_al}.

%% file: methodology.tex
\section{Methodology}
\label{sec:method}
\begin{figure}
    \centering
    \includegraphics[trim=0 0.5cm 0 0,clip,width=0.8\linewidth]{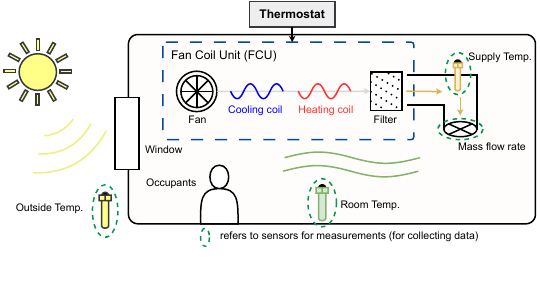}
    \vspace{-7pt}
    \caption{In-door thermal dynamics system.}
    \label{fig:HVAC}
\end{figure}

This study compares the performance of optimal experimental design (OED) based active learning, namely the fourteen methods presented in Sections~\ref{sec:landscape} through~\ref{sec:stochastic_al}, against passive learning for accurately identifying building thermal dynamics.
Specifically, the thermal dynamics of the HVAC system presented in our previous work~\cite{11346501} are considered, since space conditioning consumes approximately 50\% of the total energy used in buildings.
Figure~\ref{fig:HVAC} illustrates the HVAC dynamics considered in this paper, where four sensors collect data on the room temperature, supply temperature, outside temperature, and the mass flow rate controlled by the fan.
These collected data are used to learn the thermal dynamics and accurately predict the next room temperature from the current measurements.
Here we formulate our system as
\begin{align}
    {x}_{k+1} = f_\text{HVAC}({x}_k, \mathbf{u}_k, {d}_k)  \quad \text{or} \quad {T}^r_{k+1} = f_\text{HVAC}({T_k^r}, [{T_k^s}, \dot{m}_k], {T_k^\text{out}}) \label{eq:HVAC_ODE}
\end{align}
where the state $x := T^r$ is the room temperature, the control inputs $\mbf{u}$ cover the supply temperature $T^s$ and the mass flow rate $\dot{m}$, $d$ is the measurable disturbance representing the outside temperature $T^\text{out}$. 
Additional measurable disturbances also exist, such as radiant heat gains from outside and occupant behavior, which are not explicitly modeled as part of $d$ in this work.
Here, to specifically investigate the thermal dynamics, we omit the thermostat operation and consider directly the supply temperature and mass flow rate.

To obtain a realistic evaluation of both the active learning and passive learning methods, this study employs a high-fidelity simulator, the Building Optimization Testing Framework (BOPTEST)~\cite{Blum03092021}, which uses Modelica~\cite{Wetter04072014} to model realistic HVAC dynamics derived from real-world systems.
The \texttt{BESTEST} Case is used for the performance comparison, since it has the same structure as the indoor thermal dynamics presented in Figure~\ref{fig:HVAC}.

As presented in Section~\ref{sec:landscape}, online learning is the central technique investigated in this study.
It is worth noting that retraining the machine learning models at every sample time, immediately after each new data point is gathered, imposes a substantial computational burden.
For this reason, a batch update technique is proposed, in which the model is retrained only after $N_\text{batch}$ new data points have been collected.

Furthermore, when conducting experiments by adjusting the control inputs $\mbf{u}$, ramp constraints must be imposed to avoid inconsistent behavior of the thermostat system~\cite{8443729}.
In other words, the current control input cannot vary excessively relative to the previous control signal, in addition to satisfying the boundary control input constraints
 \begin{subequations}
 \label{eq:constraint}
\begin{align}
    & |\mbf{u}_k - \mbf{u}_{k-1}| \leq \text{ramp constraints},\\
    & \mbf{u}_\text{min} \leq \mbf{u} \leq \mbf{u}_\text{max},
\end{align}
\end{subequations}
where ramp denotes the ramp constraint, and $|\cdot|$ denotes the elementwise absolute value of the vector $\mbf{u}_k - \mbf{u}_{k-1}$.
The final algorithm used for online learning is presented in Algorithm~\ref{alg:batch}.

For general evaluation, the root mean square error (RMSE) of each model is computed at every time step over the fixed test dataset
\begin{align*}
    \text{RMSE} = \sqrt{\frac{1}{N_\text{test}} \sum_{i = 1}^{N_\text{test}} (x_{i} - \hat{x}_i)^2},
\end{align*}
where a smaller RMSE indicates better performance in identifying the thermal dynamics of the HVAC system.

To characterize how the RMSE evolves as additional experiments are collected during online learning, the RMSE is evaluated over three nested time windows measured from the start of the test horizon, namely the first two hours (0--2\,h), the first twelve hours (0--12\,h), and the full twenty-four-hour horizon (0--24\,h).
For each window, two summary statistics are reported: the windowed mean RMSE, obtained by averaging the instantaneous RMSE over all time steps within the window, and the RMSE at the final time step of the window, denoted Last, which reflects the model accuracy attained by the end of that window rather than its average accuracy throughout.

\begin{algorithm}[t]
    \caption{Batch Update for Online Learning}
    \label{alg:batch}
    \small
    \begin{algorithmic}
    \Require Initialize the training data $\mcal{D}$, initialize ML model parameters $\theta$, number of experiments $N_\text{budget}$.
    \Ensure Optimal parameters $\theta^*$ 
    \State Train $\theta_0 \leftarrow \argmin_{\theta} \mcal{L}(\theta; \mcal{D})$.
    \For{$\;\;i \!: 1 \rightarrow N_\text{budget}$}
        \State Given $\mbf{x}_i$, choose either random $\mbf{u}_i$ (passive learning) or optimal $\mbf{u}_i$ (active learning), subject to the constraints \eqref{eq:constraint}.
        \State Given $[\mbf{x}_i, \mbf{u}_i]$, we collect $\mbf{x}_{i+1}$ via \eqref{eq:HVAC_ODE}.
        \State Update the dataset $\mcal{D} = \mcal{D} \cup \{ [\mbf{x}_i, \mbf{u}_i], \mbf{x}_{i+1}\}$.
        \If{$i\;\; \text{mod} \;\; N_\text{batch} == 0$} \quad
            \State Retrain the ML model $\theta_i \leftarrow \argmin_{\theta} \mcal{L}(\theta; \mcal{D})$
        \EndIf
    \EndFor
    \State Save optimal parameters $\theta^* \leftarrow \theta_{N_\text{budget}}$.
    \end{algorithmic}
\end{algorithm}

%% file: results.tex
\section{Simulation Results}
\label{sec:results}

To evaluate the performance of the optimal experimental design techniques investigated in this study, an online learning scenario is examined in which new experiments are collected at a fixed sample time of five minutes, and the machine learning models are retrained every $N_\text{batch} = 10$ collected samples, corresponding to a batch update interval of fifty minutes, following the procedure described in Section~\ref{sec:method}.

The initial training dataset and the held-out test dataset are both collected using a uniform sampling distribution over the \texttt{peak-heat-day} scenario of the \texttt{BESTEST} Case, spanning a total simulation horizon of 14 days.
The admissible control input space is bounded by the supply temperature constraint $T^s \in [12^\circ\text{C}, 40^\circ\text{C}]$ and the normalized mass flow rate constraint $\dot{m} \in [0\%, 100\%]$.

During the online learning process, experiments are collected over the first day of the \texttt{peak-heat-day} scenario, a twenty-four-hour horizon over which the outside temperature $T^\text{out}$ varies between $-5^\circ\text{C}$ and $0^\circ\text{C}$.
The trained models are subsequently evaluated on the test set drawn from day six of the same scenario, over which $T^\text{out}$ varies between $-5^\circ\text{C}$ and $10^\circ\text{C}$.
Since the test horizon spans an outside temperature range that is largely disjoint from the training horizon, this evaluation also constitutes a mild out-of-distribution generalization test for the trained models.

For a comprehensive analysis, three test case scenarios are investigated, each pairing an initial training dataset size with a distinct ramp constraint on the rate of change of the control inputs:
\begin{itemize}
    \item Two initial data points, with ramp constraints of $0.8^\circ\text{C}$ on $T^s$ and 2\% on $\dot{m}$, and with wider ramp constraints of $2^\circ\text{C}$ on $T^s$ and 5\% on $\dot{m}$.
    \item 10 initial data points, with the widest ramp constraints of $8^\circ\text{C}$ on $T^s$ and 20\% on $\dot{m}$.
\end{itemize}
The two-point scenarios impose tight ramp constraints that limit how quickly the control inputs can change during data collection, thereby reducing the diversity of the resulting input excitation; nevertheless, such tight constraints are more representative of real-world HVAC operation, where abrupt setpoint changes would compromise occupant comfort and equipment safety.
In contrast, the ten-point scenario initializes the ML models with a larger initial dataset and admits a looser ramp constraint, allowing the control inputs greater freedom to explore the admissible input space $\mcal{U}$. 
Also, for the PL technique, the control inputs follow the uniform distribution.

The following subsections report the resulting RMSE statistics for the GP model (Section~\ref{sec:results_gp}) and the NN model (Section~\ref{sec:results_nn}), followed by a comparative discussion across both model classes (Section~\ref{sec:discussion}).
Each RMSE value is reported in degrees Celsius and is computed over three nested evaluation windows, namely the first two hours, the first twelve hours, and the full twenty-four-hour test horizon, using both the windowed mean RMSE and the RMSE at the final time step of the window (denoted Last).

\begin{table}
\centering
\small
\caption{\normalsize GP model \textbar{} Initial points: 2, Ramp constraints for $T^s$: 0.8$^\circ$C, $\dot{m}$: 2\%}
\vspace{0.2cm}
\label{tab:GP_ini2_0.02}
\begin{tabular}{llccccccccc}
\toprule
Window & Metric & PL & $\mathrm{GS}_{\mathrm{GP}}$ & $\mathrm{IDW}_{\mathrm{GP}}$ & MV & IVR & FI & SE & PI & UCB \\
\midrule
\multirow{2}{*}{0--2h} & Mean & 8.846 & 9.515 & 9.369 & 9.574 & \textbf{7.269} & 8.220 & 8.565 & 8.036 & 9.554 \\
 & Last & 7.579 & 8.667 & 7.842 & 8.799 & \textbf{4.681} & 6.083 & 7.111 & 6.272 & 8.642 \\
\midrule
\multirow{2}{*}{0--12h} & Mean & 6.651 & 7.742 & 4.551 & 4.990 & \textbf{4.338} & 4.495 & 6.278 & 5.360 & 4.898 \\
 & Last & 5.467 & 6.856 & \textbf{2.503} & 2.693 & 2.797 & 2.739 & 5.030 & 4.156 & 3.114 \\
\midrule
\multirow{2}{*}{0--24h} & Mean & 5.583 & 6.857 & 3.337 & 3.677 & \textbf{3.335} & 3.485 & 5.398 & 4.546 & 3.791 \\
 & Last & 3.527 & 4.908 & \textbf{1.974} & 2.099 & 2.062 & 2.247 & 4.282 & 3.077 & 2.162 \\
\bottomrule
\end{tabular}
\end{table}
\begin{table}
\centering
\small
\caption{\normalsize GP model \textbar{} Initial points: 2, Ramp constraints for $T^s$: 2$^\circ$C, $\dot{m}$: 5\%}
\vspace{0.2cm}
\label{tab:GP_ini2_0.05}
\begin{tabular}{llccccccccc}
\toprule
Window & Metric & PL & $\mathrm{GS}_{\mathrm{GP}}$ & $\mathrm{IDW}_{\mathrm{GP}}$ & MV & IVR & FI & SE & PI & UCB \\
\midrule
\multirow{2}{*}{0--2h} & Mean & 9.225 & 9.748 & 7.929 & 8.689 & \textbf{6.731} & 6.896 & 8.439 & 8.908 & 8.483 \\
 & Last & 8.157 & 8.895 & 4.491 & 5.300 & 4.855 & \textbf{4.074} & 6.963 & 5.778 & 4.909 \\
\midrule
\multirow{2}{*}{0--12h} & Mean & 7.161 & 7.811 & 3.916 & 4.272 & 3.743 & \textbf{3.587} & 6.138 & 3.850 & 4.094 \\
 & Last & 4.559 & 6.647 & 2.203 & \textbf{2.066} & 2.405 & 2.166 & 4.901 & 2.532 & 2.806 \\
\midrule
\multirow{2}{*}{0--24h} & Mean & 5.184 & 6.726 & 2.998 & 3.172 & 2.903 & \textbf{2.795} & 5.264 & 3.007 & 3.261 \\
 & Last & 2.518 & 4.484 & 1.981 & 1.967 & 1.955 & \textbf{1.887} & 4.143 & 1.976 & 2.155 \\
\bottomrule
\end{tabular}
\end{table}
\begin{table}
\centering
\small
\caption{\normalsize GP model \textbar{} Initial points: 10, Ramp constraints for $T^s$: 8$^\circ$C, $\dot{m}$: 20\%}
\vspace{0.2cm}
\label{tab:GP_ini10_0.2}
\begin{tabular}{llccccccccc}
\toprule
Window & Metric & PL & $\mathrm{GS}_{\mathrm{GP}}$ & $\mathrm{IDW}_{\mathrm{GP}}$ & MV & IVR & FI & SE & PI & UCB \\
\midrule
\multirow{2}{*}{0--2h} & Mean & 6.108 & 4.959 & 5.229 & 4.852 & \textbf{4.664} & 5.634 & 6.813 & 4.904 & 4.883 \\
 & Last & 4.824 & 3.988 & 3.764 & \textbf{3.518} & 3.742 & 4.089 & 6.366 & 3.743 & 3.592 \\
\midrule
\multirow{2}{*}{0--12h} & Mean & 3.791 & 3.454 & 3.302 & 3.073 & 3.438 & 3.447 & 5.662 & 3.228 & \textbf{3.066} \\
 & Last & 2.301 & 2.438 & 2.368 & \textbf{2.286} & 2.734 & 2.431 & 4.650 & 2.369 & 2.403 \\
\midrule
\multirow{2}{*}{0--24h} & Mean & 3.012 & 2.847 & 2.757 & \textbf{2.567} & 2.862 & 2.840 & 4.927 & 2.685 & 2.638 \\
 & Last & 2.232 & 2.130 & 2.135 & 1.932 & \textbf{1.928} & 2.151 & 3.983 & 2.034 & 2.118 \\
\bottomrule
\end{tabular}
\end{table}

\subsection{Gaussian Process}
\label{sec:results_gp}

The GP model is implemented using GPyTorch~\cite{3327757.3327857}, with kernel hyperparameters optimized via the resilient backpropagation (RPROP) solver~\cite{298623}.
The initial GP hyperparameters are trained for five epochs prior to the onset of online learning, and the hyperparameters are subsequently re-optimized for five additional epochs at every batch update, using a learning rate of 0.01.

\begin{figure}
    \centering
    \makebox[\linewidth]{\includegraphics[width=0.9\linewidth]{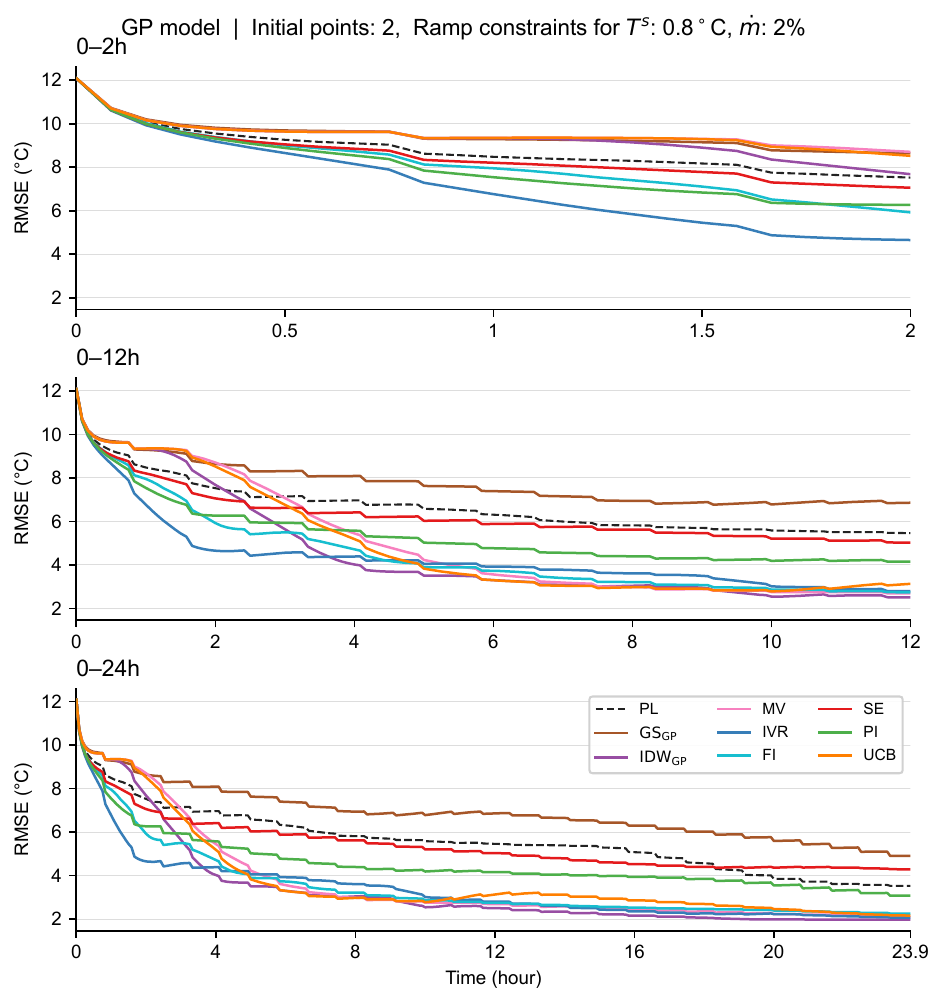}}
    \vspace{-10pt}
    \caption{Test RMSE of the GP model versus elapsed online learning time, for two initial data points and ramp constraints of $0.8^\circ\text{C}$ on $T^s$ and 2\% on $\dot{m}$, across the 0--2\,h, 0--12\,h, and 0--24\,h evaluation windows (Table~\ref{tab:GP_ini2_0.02}).}
    \label{fig:GP_ini2_0.02}
\end{figure}

\begin{figure}
    \centering
    \makebox[\textwidth]{\includegraphics[width=0.9\linewidth]{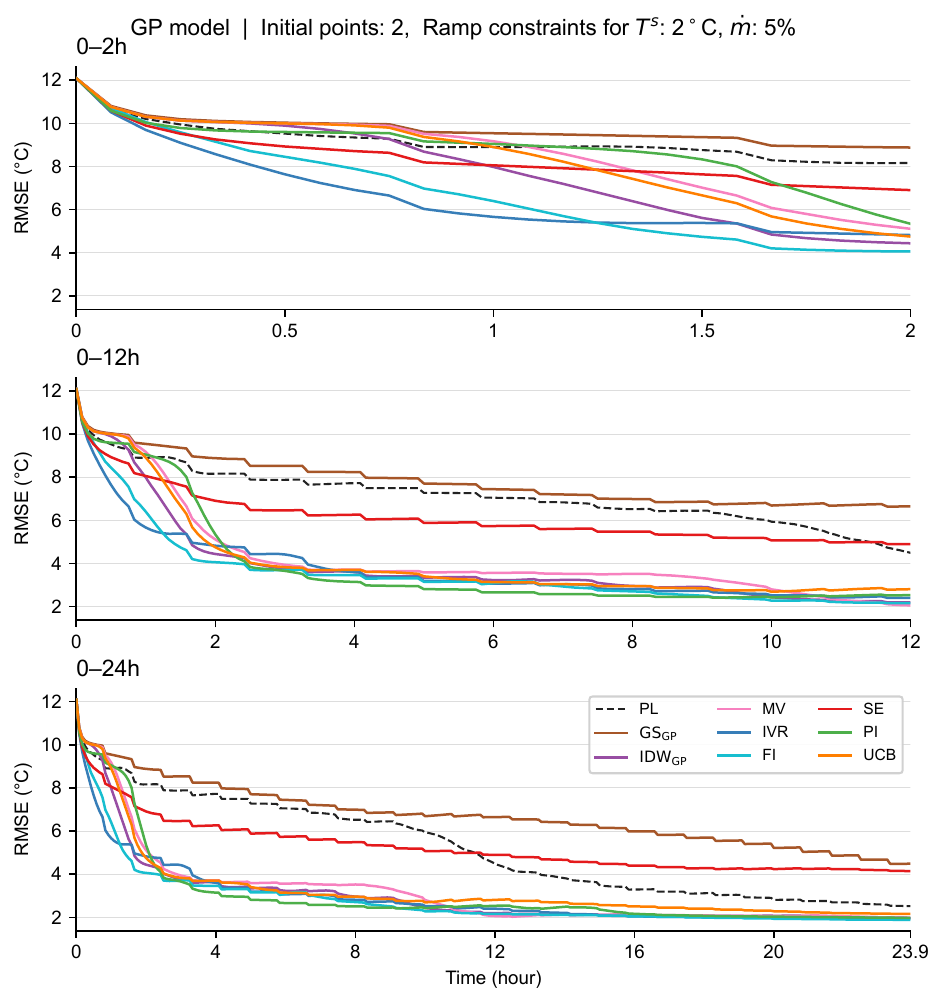}}
    \vspace{-10pt}
    \caption{Test RMSE of the GP model versus elapsed online learning time, for two initial data points and ramp constraints of $2^\circ\text{C}$ on $T^s$ and 5\% on $\dot{m}$, across the 0--2\,h, 0--12\,h, and 0--24\,h evaluation windows (Table~\ref{tab:GP_ini2_0.05}).}
    \label{fig:GP_ini2_0.05}
\end{figure}

\begin{figure}
    \centering
    \makebox[\linewidth]{\includegraphics[width=0.9\linewidth]{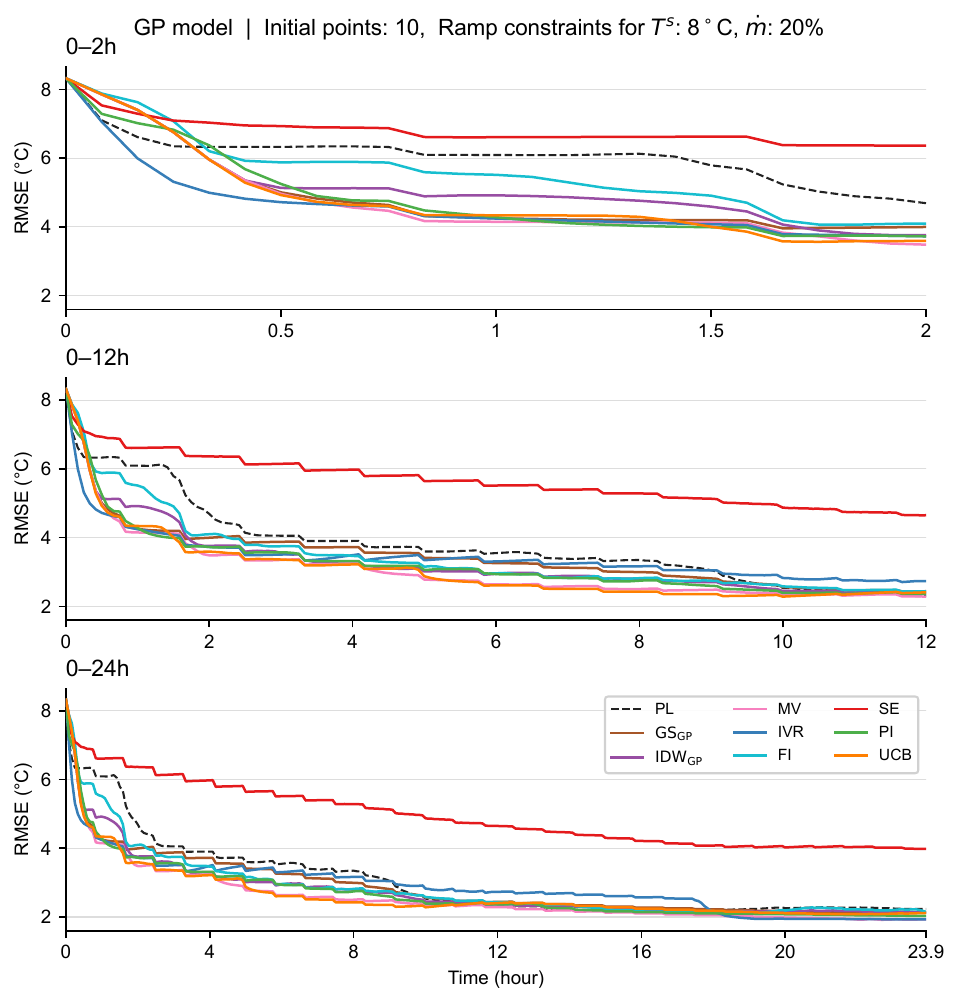}}
    \vspace{-10pt}
    \caption{Test RMSE of the GP model versus elapsed online learning time, for 10 initial data points and ramp constraints of $8^\circ\text{C}$ on $T^s$ and 20\% on $\dot{m}$, across the 0--2\,h, 0--12\,h, and 0--24\,h evaluation windows (Table~\ref{tab:GP_ini10_0.2}).}
    \label{fig:GP_ini10_0.2}
\end{figure}

With only two initial data points and the tightest ramp constraint examined in this study, IVR attains the lowest mean RMSE in both short-horizon windows, reaching $7.269^\circ\text{C}$ at the 0--2\,h window, an 18\% reduction relative to the PL baseline of $8.846^\circ\text{C}$, and $4.338^\circ\text{C}$ at the 0--12\,h window, a 35\% reduction relative to PL.
This early advantage is consistent with the global perspective of the IVR acquisition function, which targets the candidate input that maximizes the integrated variance reduction over the entire admissible space rather than only the locally most uncertain point, a property expected to matter most when very few labeled samples constrain the GP posterior.
By the end of the horizon, IDW$_\text{GP}$ attains the lowest RMSE at the final time step in both the 0--12\,h window ($2.503^\circ\text{C}$, a 54\% reduction relative to PL) and the 0--24\,h window ($1.974^\circ\text{C}$, a 44\% reduction relative to PL), reflecting the benefit of its model-based exploitation term once the GP mean prediction becomes sufficiently reliable to guide sampling.
In this scenario, GS$_\text{GP}$ is the only method that fails to improve over PL in any window or metric, attaining a higher RMSE than PL throughout the full horizon, which is consistent with its purely geometric and model-agnostic acquisition rule providing no advantage when the GP posterior itself is already informative about where additional samples are most needed.

With a looser ramp constraint, FI emerges as the strongest method over most of the horizon, attaining the lowest RMSE at the final time step of the 0--2\,h window ($4.074^\circ\text{C}$, a 50\% reduction relative to PL), the lowest mean RMSE over the 0--12\,h window ($3.587^\circ\text{C}$, a 50\% reduction relative to PL), and both the lowest mean and final RMSE over the 0--24\,h window ($2.795^\circ\text{C}$ and $1.887^\circ\text{C}$, reductions of 46\% and 25\% relative to PL, respectively).
This pattern is consistent with the theoretical role of FI in this study, since maximizing the determinant of the Fisher information matrix directly targets experiments that most tightly constrain the kernel hyperparameters, a benefit expected to be largest when hyperparameter uncertainty is still substantial early in the learning process.
GS$_\text{GP}$ again fails to improve over PL in any window, and SE trails PL in three of the six reported rows, including a narrow underperformance at the 0--24\,h mean ($5.264^\circ\text{C}$ versus $5.184^\circ\text{C}$ for PL).
PI also performs consistently well in this scenario, outperforming PL in every reported row, although it never attains the lowest RMSE in any individual window.

With 10 initial data points and the loosest ramp constraint examined in this study, the advantage of OED over PL narrows considerably relative to the two-point scenarios.
IVR and MV remain the strongest performers over the short horizon, with IVR attaining a mean RMSE of $4.664^\circ\text{C}$ at the 0--2\,h window, a 24\% reduction relative to the PL value of $6.108^\circ\text{C}$, and MV attaining the lowest final RMSE of $3.518^\circ\text{C}$ at the same window, a 27\% reduction relative to PL.
However, at the final time step of the 0--12\,h window, MV attains an RMSE of $2.286^\circ\text{C}$, less than 1\% lower than the PL value of $2.301^\circ\text{C}$, while every other method, namely GS$_\text{GP}$, IDW$_\text{GP}$, IVR, FI, SE, PI, and UCB, reports a higher RMSE than PL at this same window and metric.
This near-parity indicates that, once a moderately sized initial dataset and a wide admissible control range are available, several OED methods provide little to no measurable benefit over PL at the medium-term horizon, even though clear advantages persist at the shortest horizon.
SE remains the most consistent underperformer in this scenario as well, trailing PL in every reported row, including a final-time RMSE of $3.983^\circ\text{C}$ at the 0--24\,h window, nearly 79\% higher than the PL value of $2.232^\circ\text{C}$, suggesting that the additional hyperparameter-uncertainty term in the SE acquisition function, described in Section~\ref{sec:se}, contributes little useful signal once the kernel hyperparameters are already reasonably well constrained by 10 initial samples.

\subsection{Neural Network}
\label{sec:results_nn}

The NN model is implemented using PyTorch~\cite{paszke2019pytorch}, with parameters optimized via the ADAM solver~\cite{kingma2017adammethodstochasticoptimization}.
The network architecture with three layers consists of four input nodes, eight hidden nodes, and one output node, trained with a learning rate of 0.0005.
The initial NN parameters are trained for 500 epochs prior to the onset of online learning, and the parameters are subsequently re-optimized for 100 epochs at every batch update.
The QBC, EMCM, and MLLC methods each use a bootstrap ensemble of 10 NN models, and the MCD method uses a dropout rate of 10\% at inference time.

\begin{table}[t]
\centering
\small
\caption{\normalsize NN model \textbar{} Initial points: 2, Ramp constraints for $T^s$: 0.8$^\circ$C, $\dot{m}$: 2\%}
\vspace{0.2cm}
\label{tab:NN_ini2_0.02}
\begin{tabular}{llccccccc}
\toprule
Window & Metric & PL & $\mathrm{GS}_{\mathrm{NN}}$ & $\mathrm{IDW}_{\mathrm{NN}}$ & QBC & MCD & EMCM & MLLC \\
\midrule
\multirow{2}{*}{0--2h} & Mean & 6.403 & 6.442 & 6.553 & \textbf{5.514} & 7.059 & 5.917 & 5.917 \\
 & Last & 4.148 & 4.175 & 4.211 & 5.621 & 4.738 & \textbf{2.692} & \textbf{2.692} \\
\midrule
\multirow{2}{*}{0--12h} & Mean & 3.414 & 3.602 & 3.174 & 3.230 & 3.970 & \textbf{2.958} & \textbf{2.958} \\
 & Last & 2.432 & 3.127 & \textbf{2.107} & 2.687 & 3.629 & 2.571 & 2.571 \\
\midrule
\multirow{2}{*}{0--24h} & Mean & 2.877 & 3.351 & 2.635 & \textbf{2.598} & 3.414 & 2.729 & 2.727 \\
 & Last & 2.352 & 3.099 & 2.143 & \textbf{1.913} & 2.015 & 2.304 & 2.292 \\
\bottomrule
\end{tabular}
\end{table}

\begin{figure}
    \centering
    \makebox[\linewidth]{\includegraphics[width=0.9\linewidth]{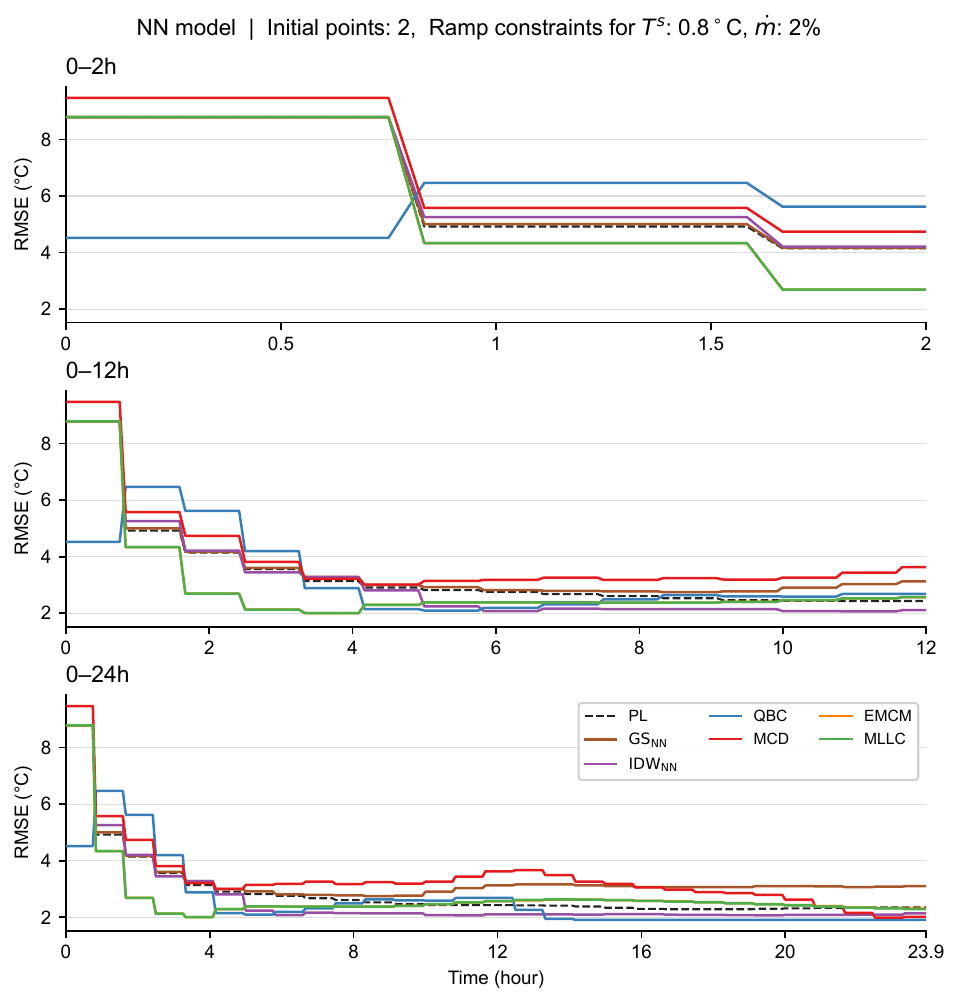}}
    \vspace{-10pt}
    \caption{Test RMSE of the NN model versus elapsed online learning time, for two initial data points and ramp constraints of $0.8^\circ\text{C}$ on $T^s$ and 2\% on $\dot{m}$ (Table~\ref{tab:NN_ini2_0.02}).}
    \label{fig:NN_ini2_0.02}
\end{figure}
\begin{figure}
    \centering
    \makebox[\linewidth]{\includegraphics[width=0.9\linewidth]{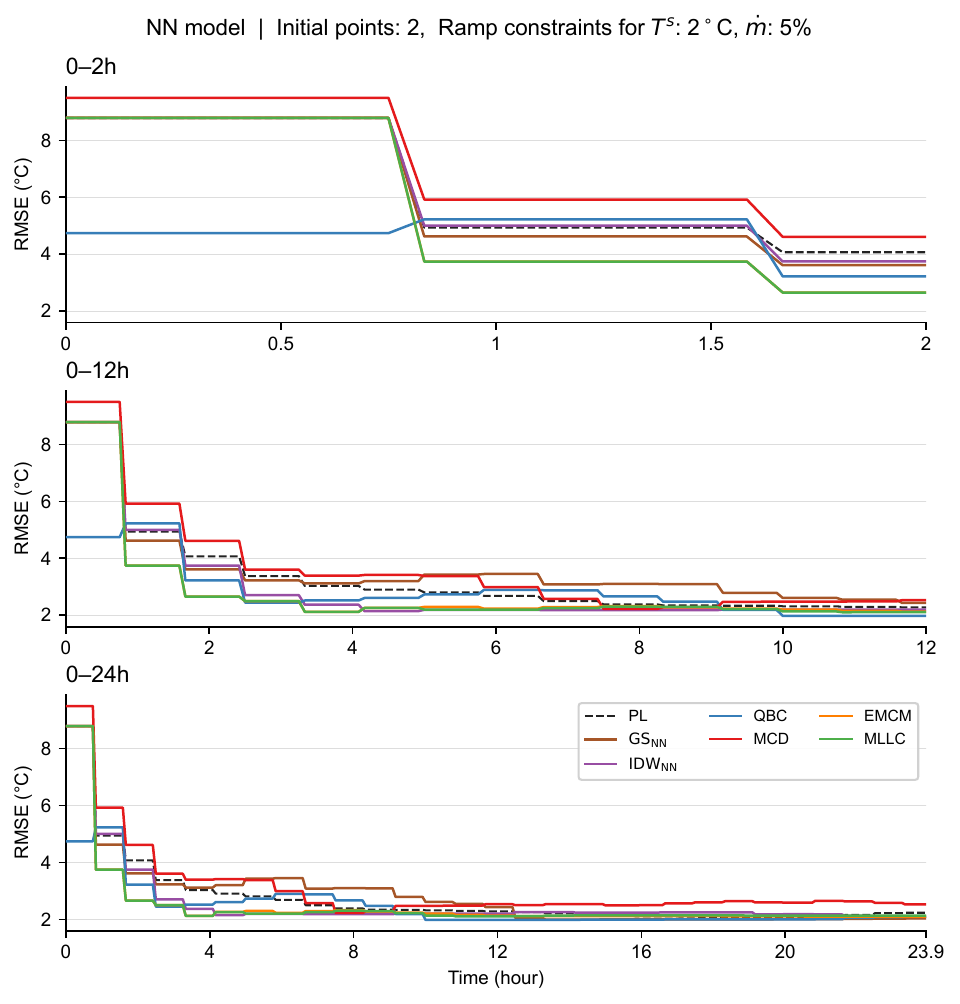}}
    \vspace{-10pt}
    \caption{Test RMSE of the NN model versus elapsed online learning time, for two initial data points and ramp constraints of $2^\circ\text{C}$ on $T^s$ and 5\% on $\dot{m}$ (Table~\ref{tab:NN_ini2_0.05}). }
    \label{fig:NN_ini2_0.05}
\end{figure}
\begin{figure}
    \centering
    \makebox[\linewidth]{\includegraphics[width=0.9\linewidth]{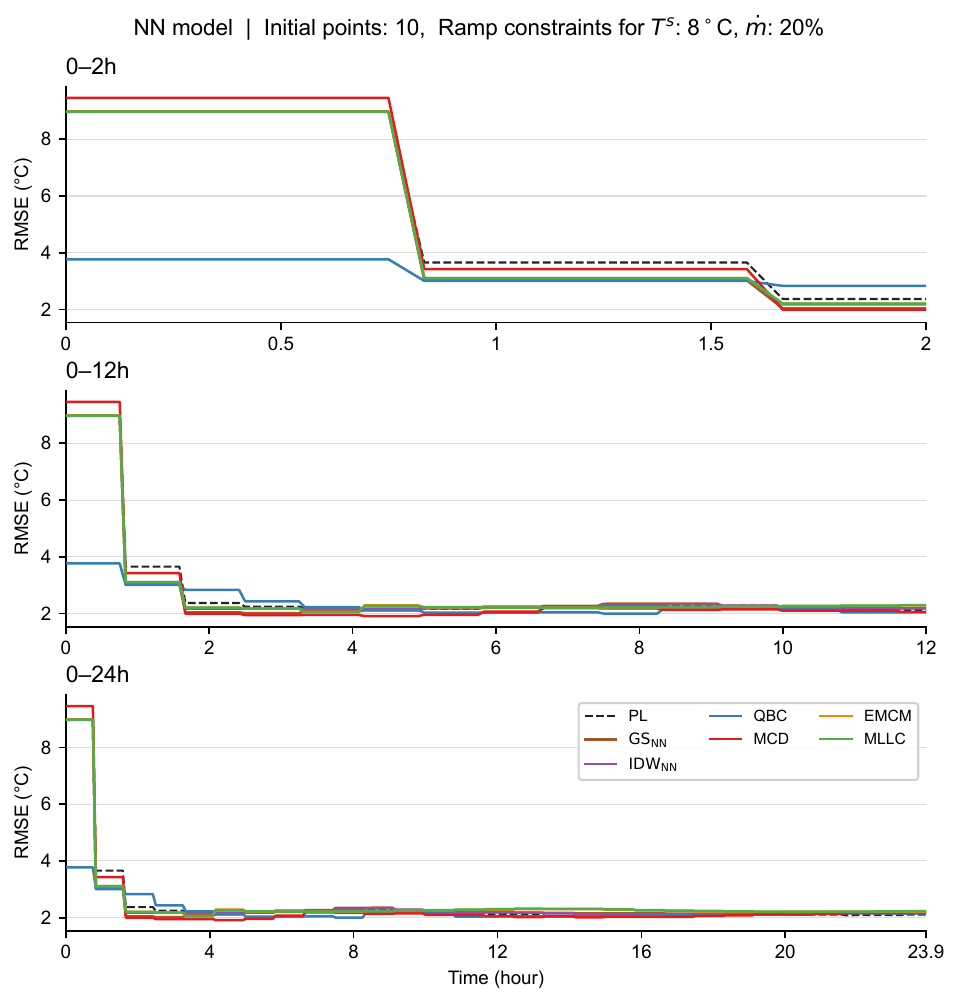}}
    \vspace{-10pt}
    \caption{Test RMSE of the NN model versus elapsed online learning time, for 10 initial data points and ramp constraints of $8^\circ\text{C}$ on $T^s$ and 20\% on $\dot{m}$ (Table~\ref{tab:NN_ini10_0.2}). }
    \label{fig:NN_ini10_0.2}
\end{figure}

\begin{table}[t]
\centering
\small
\caption{\normalsize NN model \textbar{} Initial points: 2, Ramp constraints for $T^s$: 2$^\circ$C, $\dot{m}$: 5\%}
\vspace{0.2cm}
\label{tab:NN_ini2_0.05}
\begin{tabular}{llccccccc}
\toprule
Window & Metric & PL & $\mathrm{GS}_{\mathrm{NN}}$ & $\mathrm{IDW}_{\mathrm{NN}}$ & QBC & MCD & EMCM & MLLC \\
\midrule
\multirow{2}{*}{0--2h} & Mean & 6.397 & 6.190 & 6.370 & \textbf{4.690} & 7.192 & 5.664 & 5.664 \\
 & Last & 4.071 & 3.615 & 3.746 & 3.223 & 4.613 & \textbf{2.656} & \textbf{2.656} \\
\midrule
\multirow{2}{*}{0--12h} & Mean & 3.310 & 3.586 & 2.998 & 2.871 & 3.634 & 2.842 & \textbf{2.820} \\
 & Last & 2.277 & 2.433 & 2.196 & \textbf{1.977} & 2.530 & 2.129 & 2.103 \\
\midrule
\multirow{2}{*}{0--24h} & Mean & 2.726 & 2.805 & 2.600 & \textbf{2.439} & 3.099 & 2.475 & 2.475 \\
 & Last & 2.232 & \textbf{2.032} & 2.086 & 2.096 & 2.523 & 2.099 & 2.132 \\
\bottomrule
\end{tabular}
\end{table}
\begin{table}[t]
\centering
\small
\caption{\normalsize NN model \textbar{} Initial points: 10, Ramp constraints for $T^s$: 8$^\circ$C, $\dot{m}$: 20\%}
\vspace{0.2cm}
\label{tab:NN_ini10_0.2}
\begin{tabular}{llccccccc}
\toprule
Window & Metric & PL & $\mathrm{GS}_{\mathrm{NN}}$ & $\mathrm{IDW}_{\mathrm{NN}}$ & QBC & MCD & EMCM & MLLC \\
\midrule
\multirow{2}{*}{0--2h} & Mean & 5.659 & 5.341 & 5.392 & \textbf{3.297} & 5.700 & 5.402 & 5.402 \\
 & Last & 2.373 & 2.044 & 2.179 & 2.834 & \textbf{1.999} & 2.209 & 2.209 \\
\midrule
\multirow{2}{*}{0--12h} & Mean & 2.802 & 2.719 & 2.753 & \textbf{2.358} & 2.668 & 2.750 & 2.752 \\
 & Last & 2.125 & 2.191 & 2.186 & 2.052 & \textbf{2.050} & 2.287 & 2.298 \\
\midrule
\multirow{2}{*}{0--24h} & Mean & 2.449 & 2.447 & 2.447 & \textbf{2.231} & 2.375 & 2.496 & 2.503 \\
 & Last & \textbf{2.109} & 2.209 & 2.161 & 2.150 & 2.192 & 2.211 & 2.230 \\
\bottomrule
\end{tabular}
\end{table}


For the NN model with two initial data points, QBC and the gradient-based model-change methods provide the largest improvements over PL.
QBC attains the lowest mean RMSE in both the 0--2\,h window ($4.690^\circ\text{C}$, a 27\% reduction relative to the PL value of $6.397^\circ\text{C}$) and the 0--24\,h window ($2.439^\circ\text{C}$, an 11\% reduction relative to PL), while EMCM and MLLC jointly attain the lowest final RMSE of $2.656^\circ\text{C}$ at the 0--2\,h window, a 35\% reduction relative to PL.
This early advantage of the gradient-based methods is consistent with their acquisition functions directly targeting candidate inputs expected to induce the largest update to the NN parameters, a property that is most valuable while the model is still poorly constrained by only two initial samples.
MCD is the only method that fails to improve over PL in any of the six reported rows, with a final RMSE of $2.523^\circ\text{C}$ at the 0--24\,h window compared to $2.232^\circ\text{C}$ for PL, suggesting that the dropout-based uncertainty estimate is insufficiently calibrated to guide sampling effectively in this configuration, consistent with the approximate nature of MC dropout as a Bayesian uncertainty proxy discussed in Section~\ref{sec:mcd}.

With 10 initial data points, QBC attains the lowest mean RMSE in every window, reaching $3.297^\circ\text{C}$ at the 0--2\,h window, a 42\% reduction relative to the PL value of $5.659^\circ\text{C}$, and $2.231^\circ\text{C}$ at the 0--24\,h window, a 9\% reduction relative to PL.
However, QBC itself reports a higher final RMSE than PL at the 0--2\,h window ($2.834^\circ\text{C}$ versus $2.373^\circ\text{C}$ for PL, 19\% higher), illustrating that the committee-disagreement signal can be noisy at the instantaneous level even while remaining beneficial on average over the window.
At the final time step of the 0--12\,h window, GS$_\text{NN}$, IDW$_\text{NN}$, EMCM, and MLLC all report a higher RMSE than PL, and at the final time step of the full 0--24\,h horizon, PL itself attains the lowest raw RMSE of $2.109^\circ\text{C}$ among all seven methods, ahead of the best-performing OED method, QBC, at $2.150^\circ\text{C}$.
This outcome indicates that, with a larger initial dataset and a wide admissible control range, the long-horizon advantage of active sampling over passive sampling for the NN model can vanish entirely at specific evaluation points, even though most OED methods retain a measurable advantage when averaged over each window.

\subsection{Discussion}
\label{sec:discussion}

Across the six scenarios examined for both the GP and NN models, OED-based sampling reduces RMSE relative to PL in the substantial majority of the reported window and metric combinations, with reductions reaching as high as 54\% for IDW$_\text{GP}$ at the 0--12\,h window in the tightest-ramp GP scenario and 42\% for QBC at the 0--2\,h window in the ten-point NN scenario.
The largest relative improvements consistently occur in the data-scarce regime of two initial points combined with tight ramp constraints, where the model is least informed and where the geometric exploration of data-space methods, the global variance reduction of IVR, and the gradient-based targeting of EMCM and MLLC each provide a clear and consistent benefit over uniformly random sampling.
This pattern is consistent with the central premise of OED investigated throughout this study: that informative sampling becomes most valuable precisely when the available data budget is smallest.

Nevertheless, the advantage of OED over PL is neither universal across methods nor uniform across scenarios, and several specific exceptions are worth stating plainly rather than averaging away.
Within the GP model, GS$_\text{GP}$ fails to improve over PL in any reported row of either two-point scenario, and SE trails PL in the majority of rows across all three GP scenarios, including the ten-point scenario where its final-window RMSE is nearly 79\% higher than PL.
Within the NN model, MCD fails to improve over PL in any reported row of the two-point scenario.
Furthermore, as the initial dataset grows to 10 points and the ramp constraint loosens, the residual advantage of OED narrows substantially at the medium- and long-term horizons for both model classes: in the GP model, MV improves on PL by less than 1\% at the final time step of the 0--12\,h window, with every other OED method reporting a higher RMSE than PL at that same point; in the NN model, PL itself attains the single lowest RMSE among all seven methods at the final time step of the full 0--24\,h horizon.
These results indicate that the benefit of OED-based sampling for building thermal system identification is most pronounced, and most reliably realized, when the initial training data are scarce and the admissible control range is tightly constrained, while the benefit of any individual acquisition strategy should not be assumed to persist uniformly once a moderately informative initial dataset is already available.

%% file: Conclusion.tex
\section{Conclusion}
\label{sec:conclusion}

This paper classifies active learning techniques for building energy system identification into four categories, namely data space, uncertainty, information gain, and model change, and formally derives the corresponding acquisition functions for two machine learning model classes: a deterministic feedforward neural network and a stochastic Gaussian process.
The resulting fourteen techniques are evaluated on the high-fidelity building simulator BOPTEST, where AL is shown to generally reduce the root mean square error relative to PL with uniformly random control inputs, although the magnitude and consistency of this improvement vary across acquisition functions and test scenarios.
In particular, AL applied to the Gaussian process model yields the largest relative accuracy gains observed in this study, with reductions in error of up to 54\% relative to PL in the data-scarce, tightly constrained regime.
Within the Gaussian process results, the data space and uncertainty categories provide the strongest performance in the tightest-constraint, data-scarce scenario, whereas the information gain category becomes the strongest once the ramp constraint is moderately loosened.
For the NN model, the benefit of AL is most pronounced for the model change methods during the first two hours of online learning, while uncertainty sampling, particularly query-by-committee, remains the most consistently competitive category across the full evaluation horizon; nevertheless, this benefit narrows considerably, and can occasionally vanish entirely, once a larger initial dataset and a looser ramp constraint are available.
These findings indicate that AL-based experimental design is most valuable, and most reliably realized, when the available training data are scarce and the admissible control range is tightly constrained, and this study can serve as a baseline benchmark for future work on active learning for HVAC systems.
Although this paper demonstrates the benefit of AL for building thermal system identification, it considers a relatively simple system with four model inputs and one output, and does not assess the computational cost of the investigated algorithms.
In future work, we will extend this study to more complex nonlinear building systems and incorporate statistical analysis over repeated trials, in addition to the computational cost of each acquisition function.